\documentclass[preprint]{aastex63}
\usepackage{subfigure}

\received{}
\revised{}
\accepted{}

\shortauthors{Liu et al.}

\begin{document}

\title{Comparative study of TESS photometry and radial velocities on six early K-type contact binaries with similar periods around 0.268 days}

\correspondingauthor{N.-P. Liu}
\email{E-mail: lnp@ynao.ac.cn}

\author[0000-0002-2276-6352]{N.-P. Liu}
\affiliation{Yunnan Observatories, Chinese Academy of Sciences,  \\
P.O. Box 110, Kunming 650216, P.R.China}
\affiliation{Key Laboratory for the Structure and Evolution of Celestial Objects,  \\
Chinese Academy of Sciences, P.O. Box 110, Kunming 650216, P.R.China}

\author{S.-B. Qian}
\affiliation{Yunnan Observatories, Chinese Academy of Sciences,  \\
P.O. Box 110, Kunming 650216, P.R.China}
\affiliation{Key Laboratory for the Structure and Evolution of Celestial Objects,  \\
Chinese Academy of Sciences, P.O. Box 110, Kunming 650216, P.R.China}
\affiliation{University of Chinese Academy of Sciences, \\
Yuquan Road 19$^{\#}$, Shijingshan Block, Beijing 100049, P.R.China}

\author{W.-P. Liao}
\affiliation{Yunnan Observatories, Chinese Academy of Sciences,  \\
P.O. Box 110, Kunming 650216, P.R.China}
\affiliation{Key Laboratory for the Structure and Evolution of Celestial Objects,  \\
Chinese Academy of Sciences, P.O. Box 110, Kunming 650216, P.R.China}
\affiliation{University of Chinese Academy of Sciences, \\
Yuquan Road 19$^{\#}$, Shijingshan Block, Beijing 100049, P.R.China}

\author{Y. Huang}
\affiliation{South-Western Institute for Astronomy Research,   \\
Yunnan University, Kunming 650500, P.R.China}

\author{Z.-L. Yuan}
\affiliation{Department of Physics, School of Physics and Electronics, Hunan Normal University,  \\
P.O. Box 110, Changsha 410081, P.R.China}

\begin{abstract}

High-precision light curves were extracted from TESScut images. Together with APOGEE and LAMOST medium resolution spectra, a joint study was made for six early K-type contact binary candidates selected unbiasedly with orbital periods around 0.268 days. It is found that all of them (RV CVn, EK Com, V384 Ser, V1038 Her, EH CVn, and CSS$\_$J125403.7+503945) are W-subtype shallow contact systems though with different mass ratios ($1/q$ = 0.27--0.62). The effective temperature differences between binary components are around a few hundred Kelvins. The original definition of A- and W-subtypes were compared with the customarily used methods which rely on the shape or on the photometric solutions of light curves. The latter two methods are not always reliable and therefore the radial velocity analysis is strongly recommended. Through a collection of all available K-type contact binaries with both photometric and spectroscopic measurements, it is found that almost all of them are W-subtype systems, except a few objects which have nearly identical temperatures for binary components. This W-subtype phenomenon for K-type contact binaries should be further checked with more samples in the future. Finally, the physical parameters of the targets were determined with joint data analysis and the multiplicity is discussed for these targets. V384 Ser and RV CVn are confirmed very likely to be triple systems from comprehensive analysis, while V1038 Her is a candidate of a triple system based on photometric and spectroscopic solutions.

\end{abstract}

\keywords{binaries: eclipsing -- stars: late-type -- stars: 
fundamental parameters -- stars: individual: RV CVn, EK Com, V384 Ser, V1038 Her, EH CVn, CRTS J125403.7+503945}

\section{Introduction}\label{secintro}
According to the definition introduced by \citet{Binnendijk70}, {late-type (or W UMa type) contact binaries (CBs) are classified into two classes: A type and W type. We use ``subtype" instead of ``type" as some new literature used \citep{LiK20, LiK21, Qian20, Lul20, Alton21, Panchal21}. For A-subtype contact systems, the more massive component has {a higher effective temperature {compared to the less massive component, while for W-subtype systems, the less massive component has {the higher temperature \citep{Yildiz13,LiK21}. This definition is more clearly illustrated with radial velocity (RV) variations of real samples. For a W-subtype system (e.g. W UMa), after the primary eclipse (the eclipse when the hotter component is eclipsed), the more massive component is receding from us (red shifted) while the less massive component is moving towards us (blue shifted). Through the analysis of RV curves, it is easy to distinguish which component is the more massive one, and thus obtain the subtype configuration. 

For totally eclipsing (TE) systems, the definition of A/W classification can be translated into the following criterion based upon the features of light curves. In A-subtype systems, the deeper minimum is caused by the transit, while in W-subtype systems, it is caused by the occultation of the less massive component \citep{Latkovic21b}. However, not all CBs are TE systems. Partially eclipsing (PE) systems are more common. Meanwhile, even for TE systems, the task of classifying the A- and W-subtypes {becomes challenging when the light curves have comparable depth at the transit and occultation phases. A good example is 1SWASP J064501.21+342154.9 (hereafter J0645). \cite{LiuN14a} classified this system to be W-subtype according to the photometric solutions based on the original definition. Then, a later study \citep{Djurasevic16} re-classified it to be A-subtype relying only on the transit-occultation feature of the light curves, but in fact, their photometric solutions also support the W-subtype configuration. This ambiguity {in classifying the subtypes occurs because the transit and occultation of the light curves have almost identical depths. The slightly shallow occultation is actually the eclipse of the hotter component. {A similar feature is shown in the case of TYC 01664-0110-1 \citep{Alton16}.

On the other side, there are more PE systems that cannot be classified using the criterion above. Photometric solution is therefore the fundamental tool. However, it is just the PE systems whose photometric solutions may have large errors \citep{Hamblek13,Liul21, LiK21}. Their mass ratios are even thought to be unreliable \citep{Hamblek13,Latkovic21b}, especially for those asymmetric light curves (in other words, spotted systems). The mass ratios derived from spectroscopic studies may be different from those based on photometric analysis (e.g. \citealt{Yakut05,Rucinski00}). Therefore, RV measurements are very important for {determining the accurate physical parameters and for {verifying the reliability of the photometric approach. However, there were only a few {late-type CBs (especially those later than G-type) studied with RV data (e.g. i Boo: \citealt{Hill89a}, SW Lac: \citealt{Zhai89}) until an RV program from David Dunlap Observatory (DDO) in Toronto (hereafter DDO program) started. This program lasted for almost ten years and a series of papers were published (e.g. \citealt{Lu99,Rucinski02,Pribulla09a}), which has greatly increased the samples of late-type CBs with RV data. Among those CBs, the K-type systems have the shortest periods which are close to the period cut-off ($\sim$ 0.22 d) according to the period-color relation \citep{Eggen67, Rucinski02}. From their RV data, it is inferred that almost all of the K-type CBs are W-subtype systems (see Section \ref{Secsum}). This result is quite different from the overall statistical result of the late-type CBs, most of which are from photometric {studies, that no significant difference in numbers is found for A- and W-subtype. \cite{Yildiz13} reported that among 100 well studied CBs, there are 51 A-subtypes and 49 W-subtypes. Recent statistical studies of CBs also show the similar result (e.g. \citealt{LiK21,Latkovic21b,Sun20}). Even for CBs with ultra-short periods, a recent statistic work finds nearly equal number of W- and A-subtype systems \citep{LiK19}.

So far, the A- and W-subtype classification has been widely used in the study of late-type CBs, not only in the statistic research (e.g. \citealt{Sun20,Latkovic21b}), but also for the formation and evolution science (e.g. \citealt{Yildiz13,Yildiz14}). Therefore, it is important to get {the correct subtype configuration. However, the difference between the spectroscopic and photometric statistic results in the W- and A-subtypes are noticeable, particularly in K-type targets. The problem is that the samples of K-type CBs with RV data are not numerous enough. Thanks to the era of  large surveys, we could collect data from both photometric and spectroscopic surveys, including the foregoing Kepler mission \citep{Borucki10,Koch10} and the present TESS (Transiting Exoplanet Survey Satellite; \citealt{Ricker15}) for the photometric part and SDSS (Sloan Digital Sky Survey: \citealt{Eisenstein11,Abolfathi18}) and LAMOST (Large Sky Area Multi-Object Fiber Spectroscopic Telescope, e.g. \cite{CuiX12,Zhao12}) for the spectroscopic part. In the present study, we mainly utilized data from TESS and APOGEE (the Apache Point Observatory Galactic Evolution Experiment; \citealt{Majewski17}).

TESS is currently one of the most important all-sky {projects ongoing, which plans initially to measure the brightness of $\sim$200,000$-$400,000 selected stars at {two-minute cadence, and simultaneously the full-frame images (FFIs) are obtained at ({30-minute cadence \citep{Stassun18}. It conducts a sequential set of 27.4 days photometric surveys of $24^{\circ}\times~96^{\circ}$ sky sectors and results in observations of 80-85$\%$ of the sky \citep{Feinstein19, Nardiello19} every two years, with a photometric precision {of about 60 ppm to 3 percent for stars brighter than $T_{mag} \approx 16$ \citep{Oelkers18}. This project is intentionally designed for exoplanet detection, which however, extremely facilitates the research of eclipsing binaries \citep[e.g.][]{Lee19,Borkovits20}. Except for the light curves from short cadence (two minutes or one minute exposure) for selected targets, FFIs of TESS provide a huge data mining archive for scientific research of diverse objects. The light curves used in this study are mostly from this archive.

APOGEE is a large near-infrared (NIR) spectroscopic survey launched within SDSS-III \citep{Eisenstein11} and later continued in SDSS-IV \citep{Blanton17}. APOGEE has produced a large collection of high resolution (R=22500), high quality (S/N $>100$), {and infrared (H-band) spectra for stars throughout the Milky Way \citep{Majewski16}. APOGEE affords us spectra of more than half a million spectra of some 150,000 stars (APOGEE-1), which is an ideal resource for spectroscopic study. However, because of its NIR band and the spectra line entanglement problem for close binaries, this treasure was not paid enough attention for the research of short-period close binaries, until recently. \citet{Jayasinghe20} combined APOGEE with other databases to study CBs. \citet{Cunningham19} used APOGEE spectra to measure RVs of close binaries with long periods, which they applied a broadening function (BF) technique \citep{Rucinski02} to accomplish. By utilizing the APOGEE spectra together with the high precision space TESS photometric data, it is hopeful to make joint studies of  K-type CBs and to provide evidence for the A/W subtype problem. 

This paper is structured as follows. The light curve extraction and data analysis are introduced in Section \ref{Secmed}. The results for individual targets are presented in Section \ref{Secrest}. We summarized and discussed our results in Section \ref{Secsum}.

\section{Data method}\label{Secmed}
\subsection{Target selection}
We tried to select a few early K-type CBs through cross-matching {the APOGEE catalogue with the VSX database \citep{Watson06}. Candidates were then checked using the TESS images (within cycle 2 of the TESS mission). The final targets were selected based on the following criteria:\\
(1) periods are within the range of 0.26--0.27 days; \\
(2) not too faint: magnitude in the V band is brighter than 14.5;\\
(3) variation amplitude is not too low: amplitudes in V or R band are larger than 0.08 mag;\\
(4) have enough spectroscopic observations: at least 3 to 5 visit spectra. If not enough, try to collect additional observations from LAMOST survey;\\
(5) do have {available observations with TESS;\\
(6) no bright companions in the close field within a distance of 5 TESS pixels (1 pixel = 21$^{\prime\prime}$). \\
These criteria ensure effective data for the present study and meanwhile, they should not have {a selection effect on the subtype problem. It is to mention that for the first criterion, the motivation is that too short periods will cause a more serious smear effect. And 0.26--0.27 days are approximately the longest periods for K-type CBs. Finally, we collected six targets selected in this study: RV CVn, EK Com, V384 Ser in group I and V1038 Her, EH CVn, and CSS$\_$J125403.7+503945 (hereafter J125403) in group II. The group labels are only for convenience. Their basic information is listed in Table \ref{tabinfo}.

\subsection{Light curves from TESS}
Our targets were observed by TESS in cycle 2 period, which is the first year of its North-sky phase. The images were taken in {30-minute cadence. The TESS FFI-cutouts (TESScut\footnote{https://mast.stsci.edu/tesscut/}: \citealt{Brasseur19}) for each target were downloaded from the service of Mikulski Archive for Space Telescopes (MAST)\footnote{https://mast.stsci.edu/portal/Mashup/Clients/Mast/Portal.html} with the help of lightkurve package (Lightkurve Collaboration 2018). The cutouts {have sizes of 10$\times$10 to 13$\times$13 pixels, depending on each target. Similar to \citet{Guo20}, the apertures and background masks from a master image were created using a percentile thresholding method and were used to generate light curves. Usually, for the raw light curves, we need to remove the {low-frequency trend using a {polynomial fitting method and remove the data with {anomalies (various spacecraft {anomalies, thermal transient characteristics, differential velocity aberration, artifacts, etc.). The final master images and the apertures of targets group I are shown in Figure \ref{Figaper}, together with the extracted light curve of RV CVn (without detrending) as an example.

\subsection{Photometric solutions}
The TESS light curves of the targets come out to be of good quality, with small errors (mostly 0.001--0.002, occasionally 0.003--0.004 mag) after removing a few outliers. Good precision is favourable for determining accurate photometric solutions \citep{Liul21}. To get the photometric solutions, the light curves were analyzed by using the {Wilson-Devinney code (hereafter the W-D code; \citealt{Wilson71,Wilson79, Wilson90,Wilson94, VanH07,Wilson08,Wilson10,Wilson12}) in the version 2013. This version not only enables the automatic calculation of limb-darkening coefficients but can also deal with light curves with {long-time exposure which may cause notable smear effects. This is very useful for the long cadence TESScut light curves of short-period CBs (typically $\sim 0.3$ days). Therefore, we incorporated the W-D code into {Python scripts in order to make it more convenient to work, especially in the grid-search process (e.g. the q-search).

In order to make a comparative study, photometric solutions were derived {independently of RVs initially. The calculation was carried out following a common procedure \citep[e.g.][]{LiuN20}. The temperature of one component was fixed while that of the other component was set free. The solutions were searched at a series of fixed mass ratios ($q=m_2/m_1$), which is a so-called q-search method \citep[e.g.][]{Qian05}. Before calculation, mode 3 (contact model) was assumed initially. The gravity-darkening coefficients were set as $g_1=g_2=0.32$ \citep{Lucy67} and the bolometric albedo was set $A_1=A_2=0.5$ \citep{Rucinski69} according to their low temperature. The square-root functions (LD=$-3$) were chosen for the treatment of limb-darkening. The limb-darkening coefficients in the band of ``TESS'' were used (the coefficients files were kindly provided by Professor Van Hamme) and interpolated in the W-D code. NGA (the number of abscissae) was set to 3 to account for the smear effect \citep{Wilson12,Zola17}). After the q-search was done, the achieved mass ratio was used as an initial input for comprehensive solutions \citep{LiuN20}. In the final calculation, the parameter of the third light should be set free, because for TESS light curves there might be blending {of light from nearby stars \citep{Guo20}. The q-search diagrams are shown in the left panels of Fig. \ref{Figlcs1} and \ref{Figlcs2}, while the observed and theoretical light curves are displayed in the corresponding right panels. The photometric solutions are shown in Tables \ref{tabwd1} and \ref{tabwd2}. {The errors listed in the parentheses are the standard deviations directly calculated by the W-D code, which, for the nonlinear situations, are not ideally correct. The real uncertainties may be three or five times larger, depending on the confidence level \citep{LiuN15}. The main characteristics of each set of photometric solutions are labeled at the head of the corresponding column. For asymmetric light curves, cool or hot spots were modeled in the calculation. Parameters of spots include the latitude $\theta$, the longitude $\psi$, the radius $r$, and the temperature factor $T_s/T_{\ast}$. They are also listed in the tables together with the spot location (on which star). Some detailed explanations of calculation and analysis for individual targets are presented in Section \ref{Secrest}.

\subsection{Radial velocity studies}
Photometric solutions derived merely from light curves may not always be reliable (see Section \ref{secintro}). To check the photometric results, we utilized the spectra from APOGEE \citep{Majewski17} as well as a few spectra from LAMOST medium-resolution survey (MRS) \citep{LiuC20,WangR19,LiuN19}. The MRS spectra cover two wavelength bands: 495--535 nm and 630--680 nm for the so-called blue band (B band) and red band (R band), respectively \citep{Zong18,WangR19}. In this data set, they are usually observed for 3 continuous short exposures. In order to avoid the H-alpha region (activity problem, see, e.g. \citealt{LiuN20}), only spectra in {the B band were used for RV determination. For APOGEE spectra, the data used in this paper were retrieved from SDSS DR14 and DR16. {The latest description {of APOGEE data can be found in \citet{Jonsson20}. Some information {on targets retrieved from APOGEE and LAMOST databases are listed in {Tables \ref{tabsp1} and \ref{tabsp2} respectively.

We {implemented the traditional cross-correlation function (CCF) method \citep{Matijevic10, Szalai07} to measure the RVs. For the APOGEE spectra, the data utilized are the individual visit spectra (apVisit). To calculate the CCFs, the sharp features (mostly fake emission and absorption) in narrow windows (usually less than 2 or 3 pixels) were first removed to get the clean spectra. They were then normalized by fitting the continuum using a Chebyshev polynomial. The normalized spectra of RV CVn from SDSS DR14 are displayed in Figure \ref{Figspn} as an example. CCFs were calculated by {cross-correlating the normalized spectra with a standard template (NGC 5272 17) which was selected from the new catalog of RV standard stars \citep{Huang18}. The CCF images are displayed in {Figures \ref{Figccf1}, \ref{Figccf2}, and \ref{Figccf3}, and the results are shown in Tables \ref{tabsp1}, \ref{tabsp2}, and \ref{tabsp3}. Finally, the RV data were fitted (Fig. \ref{Figrv}) using the W-D code, taking account of the eclipse and proximity correction \citep{Baran04, Wilson08} for binary components. The value of the semi-major axis was set free during the calculation and the absolute physical parameters were derived accordingly. The results are listed in Table \ref{tabcollect}. More details about the RV calculation are addressed in the following section.

\section{Results}\label{Secrest}

\subsection{RV CVn}
{The variability of RV CVn has been known for almost a century since its first discovery by \citet{Larink21}. It has been subsequently studied by several authors (e.g. \citealt{Schilt27, Hoffmann81, LiuN14, Zasche14}) and two sets of photometric solutions were published recently. However, their solutions are quite different. The mass ratios from \citet{LiuN14} and \citet{Zasche14} are 1.74 ($1/q=0.575$) and 0.93, respectively, which results in conclusions of W-subtype and A-subtype, respectively. Therefore, it is important to have another check from new data.

To derive independent photometric solutions for this target, {the temperature of star 1 (the star eclipsed at the deeper minimum) was estimated and fixed at 4850 K according to the recent catalogue of Gaia DR2 \citep{Gaia18}. A third light was included in the solutions because there is a close companion in its neighbourhood. It is found that RV CVn is a W-subtype contact binary with a mass ratio about 1.97. The solutions are similar to that {obtained by \citet{LiuN14} in temperature difference ($T_1-T_2 \sim 140$ K) and the degree of contact ($f = 10\%$), while there {is slight difference in the mass ratios and inclination values. Subsequently, we tried using both mass ratio values to fit the RV data and found both of them show good fits (see the upper left panel of Fig. \ref{Figrv}). The curves from the result of \citet{LiuN14} even match the data slightly better. Therefore, a more reasonable mass ratio should be $1/q = 0.55_{-0.04}^{+0.02}$. Nevertheless, the absolute physical parameters were estimated (Table \ref{tababs}) using the mass ratio result newly determined in order to make {a comparison here. The result of total luminosity ($L_{tot}$) turns out to be in good agreement with that calculated from the parallax in Gaia DR2 ($L_{G2}$, see Table \ref{tababs}). The formulae can be found in \citet{Chen18}, which were utilized to calculate $L_{G2}$.

\subsection{EK Com}
It is noticed that K-type CBs are more likely to be W-subtype, which means that the component eclipsed at the deeper minimum (namely star 1) is the hotter but less massive component, thus star 2 should be more massive and larger. It is probable that star 2 (the primary) is more luminous. Therefore, we set the temperature $T_2$ fixed henceforth.

EK Com is an interesting object which exhibits long-term variation in its light curve. Its O'Connell effect changes greatly over a long {time \citep{Tavakkoli17}. The light curves from \citet{Samec96} and \citet{Deb10} also vary in the depths of minimum light. {Those from the former show {a typical W-subtype feature (deeper occultation) while the latter show more shallow occultation. As mentioned in Section \ref{secintro}, the latter is {usually considered {to be an A-subtype LC feature. However, the photometric solution from \citet{Deb10} indicates the system to be W-subtype actually. Therefore, all photometric solutions till now agree {on the W-subtype configuration of the system, while the mass ratio value varies from 2.89 ($1/q=0.346$; \citealt{Deb10}) to 3.5 \citep{Tavakkoli17}. No spectroscopic mass ratio is obtained yet.

According to LAMOST DR7, EK Com has a spectral type G9/G8. To derive the photometric solution, $T_2$ was estimated to be 5000 K based on LAMOST and Gaia DR2. There is a small asymmetry (O'Connell effect) in the light curve (see the middle panel of Figure \ref{Figlcs1}). Therefore, a starspot was included when searching for the final solutions. We tried both cool and hot spot models, and the solutions are shown in Table \ref{tabwd1}. Both cases turned out to give good fitting results. The solutions with a hot spot has a smaller residual which means it is more reasonable to be adopted. The calculated light curves with {the smear effect removed are also displayed in Figure \ref{Figlcs1}. The flat occultation feature of the light curve is excellently recovered.

On the spectroscopic part, except for APOGEE, we also got spectra from Thai National Observatory (TNO) for EK Com. The observations were made in 2017 with the Middle Resolution ($R \sim 18000$) fiber-fed Echelle Spectrograph (MRES) mounted on the 2.4-m Ritchey-Chretien telescope of TNO, which is operated by National Astronomical Research Institute of Thailand (NARIT). Some information is listed in Table \ref{tabsp3}, together with the determined RVs. To test the photometric solutions, parameters from both cool and hot spot solutions were used to fit the collected RV data. It is found that both inputs lead to good fits (see the middle left panel of Figure \ref{Figrv}). From the RV fit, both sets of solutions should be acceptable. Both of them agree that the system is a W-subtype contact binary with a small mass ratio ($1/q \sim 0.3$) and a high inclination ($i\gtrsim 85$), which is in {accord with the results from literature \citep[see][]{Samec96,Tavakkoli17}. It should be noted that different photometric solutions indicate that the light curve alone is not yet able to well constrain the spot configuration. The absolute dimensions were derived and listed in Table \ref{tababs} (orbital parameters from the hot spot model were used).

\subsection{V384 Ser}
V384 Ser has a spectral type K2 according to LAMOST DR7. So far, there are two sets of independent photometric solutions which were reported just recently \citep{Michaels19,ZhangL20}. The system was found to be W-subtype, with mass ratios $q = 2.65$ and 3.16 according to \citet{Michaels19} and \citet{ZhangL20}, respectively. Michaels et al. found significant luminosity contribution from a third body while Zhang et al. did not. The former authors further estimated the mass of the tertiary component from the orbital period analysis and speculated its spectral type to be K3 or K4.

V384 Ser was observed by TESS in sectors 24 and 25 (hereafter s24 and s25). This object was not only observed with FFI images but also recorded in short cadence mode. However, to be consistent with other targets, the long cadence light curves (LLCs) were analyzed first. For the light curve from s24, no spots were included in the calculation since there is no obvious asymmetry, while for s25, the spots were considered because the asymmetry is significant. A significant third light was found, which contributes to $\sim10~\%$ in {the TESS band. The mass ratios determined from LLCs of the two sectors are quite different, while both of them are smaller ($1/q$ is larger) than {those got by the previous authors. 

The time resolution of short cadence light curves (SLCs) of V384 Ser is much higher and thus benefits the analysis of the long-term variation of the light curve. The SLC data of V384 Ser used in this paper can be found in MAST: \dataset[10.17909/fpt5-8n40]{http://dx.doi.org/10.17909/fpt5-8n40}. In Figure \ref{Figlcsc} (the upper left panel), we could see the change of $-\Delta Max$ and $-\Delta Min$ of the light curves is significant and fast. Cycle to cycle variation of the light curve is probably true. Four segments of data selected (labeled as ``s24a'', ``s24b'', ``s25a'' and ``s25b'') are shown in the lower left panel. They were each binned {into 300 data points. It is seen that even in the same sector, the light curves are quite different. The fast variation indicates that V384 Ser is an active system, which is the same conclusion {obtained by \citet{Michaels19}. Photometric solutions were carried out for ``s24a'' and ``s24b'' as typical examples (light curves with roughly equal {maxima and different {maxima, respectively), and the results are shown in Table \ref{tabwd4} and the upper right panel of Figure \ref{Figlcsc}. Similar to the situation of LLCs of s24 and s25, we got quite different mass ratios for ``s24a'' (no spots) and ``s24b''(with spots). When the spotted scenario was considered for ``s24a'' (very slight asymmetry in the light curve), the mass ratio turned out to be close to that of ``s24b''. 

To solve the problem of different mass ratios, a simultaneous fitting was carried out for the light curve of ``s24a'' and the RV data. These RVs were determined from APOGEE and LAMOST MRS spectra (see Tables \ref{tabsp1} and \ref{tabsp2}). The derived mass ratio from the simultaneous solutions (the ``s24a with RV'' column of Table \ref{tabwd4}) is $q\sim~1.74$. Interestingly, this value is very close to the mean result ($q=1.757$) from the other three solutions. The derived RV curves are shown in Figure \ref{Figrv}. The curves from the ``el3'' solutions of s24 ($q=1.993$) are also plotted for comparison. It is seen that the curves from the simultaneous solutions fit the observation better. Therefore, the corresponding results from ``s24a with RV'' were adopted, and the absolute dimensions were estimated accordingly.

It is to mention that the spectra of V384 Ser were found full of narrow lines overlapped on the broad absorption features (see Fig. \ref{Figspa}). These narrow lines may come from a tertiary component with slower rotation. To our expectation, prominent features of a third component were found in the CCFs (see Fig. \ref{Figccf1}), and this is in accordance with the photometric solutions. The determined RVs of the third component were close to that of the binary system (see Table \ref{tababs}). Further more, through $O-C$ analysis, the orbital period of V384 Ser is found to show periodic variation \citep{Michaels19}. Combining all these factors, V384 Ser is most likely to be a hierarchical triple system. For the luminosity contribution from the tertiary component, our estimation is smaller than Michaels's decision. In fact, the parameters from our solutions are closer to that from the solution 2 of \citet{Michaels19}, except for the mass ratio. This is understandable since the third light can make the mass ratio more uncertain \citep{Hamblek13, Liul21}. Nevertheless, the spectroscopic analysis confirms the results from our TESS photometry. Combining photometric solutions and the spectroscopic analysis, the luminosity of the tertiary component was roughly estimated to be 0.058 L$_{\sun}$, which corresponds to $\sim$ 0.5 M$_{\sun}$ (estimated from \citealt{Eker18} and \citealt{Mann19}) for a main sequence star. If the mass function from \citet{Michaels19} was well determined, the orbital inclination of the companion should be significantly higher than 45$^{\circ}$($i_3\sim 70^{\circ}$).

\subsection{V1038 Her}
Except for a few observed minimum light times \cite[e.g.][]{Brat11,Karam16,Hubscher17}, V1038 Her has no photometric solutions published yet. 
The spectral type from LAMOST LRS is K3. Through TESS photometry, the photometric solutions were derived for the first time. A third light contributing to $\sim 6.5~\%$ in {the TESS band was found and the mass ratio was found larger ($1/q~\sim$ 0.62 ) than other targets. Except for the TESS light curve, we also got light curves from American Association of Variable Star Observers (AAVSO) in B, V, and Ic bands (see Figure \ref{figlcs_V1038}) which was observed by Kevin Alton with a QSI 683 wsg-8 CCD camera at the {privately owned UnderOak Observatory. According to Kevin Alton, the exposure time for B- and Ic passbands was 120 sec while for V was 105 sec. The photometric solutions were derived independently, following the same process. The results are listed in Table \ref{tabwd3} and the calculated light curves are shown in Figure \ref{figlcs_V1038}. The solutions from TESS and AAVSO light curves are generally in good agreement.

The RVs were fitted with both parameters from the two sets of solutions. No significant difference was seen {between the two fittings (see Figure \ref{Figrv}). Therefore, the reasonable parameters for V1038 Her to be adopted should be $q = 0.634\pm 0.014$ and $i = 76.1\pm 0.6$. The temperature difference is clearly different between the two sets of solutions, which may indicate that this target is active. However, this is not conclusive until there are more observations and evidence.

\subsection{EH CVn}
The first photometric solutions of EH CVn were derived by \cite{Xiaq18}, who found the target to be a W-subtype shallow contact binary based on the ground-based observation data. They obtained a small mass ratio about $1/q \sim$ 0.30. No radial velocities were published.

We tried to revisit EH CVn with TESS photometry and RV study. The result was interesting that the mass ratios from {the q-search and free calculation process were quite different. The q-search calculation led to a mass ratio about 3.4 which is very similar to that obtained by \cite{Xiaq18}, while the free calculation got a much smaller mass ratio $q$. The spots were added in the final solutions for the asymmetry in the light curve. The final solutions listed in Table \ref{tabwd2} were calculated from two cases. For case 1, the latitude of {the spot is not  constrained, which got a pole spot solution. For case 2, the latitude of {the spot is constrained in the low latitude region. Both solutions led to good fits for the light curve (see Figure \ref{Figlcs2}), of which the mass ratios are significantly different.

The RVs would provide independent evidence for the mass ratio. It comes out that only with the parameters from case 2 can a good fit be obtained. Therefore, the large mass ratio ($1/q = 1.383$) solution was abandoned and the intermediate mass ratio was adopted for the target. For other parameters, our solutions are similar to that of \cite{Xiaq18}. No significant third light was detected in spectra or photometric solutions.

\subsection{J125403}
J125403 (cross id. CRTS J125403.7+503945) was first found to be a periodic variable by \cite{Drake14}. The EW-type like light curve together with a period about 0.27 days indicates it to be a contact binary candidate. To clarify the nature of this newly discovered eclipsing binary, photometric solutions were searched for the light curves from TESS sectors 15 and 16 (hereafter s15 and s16). However, the low amplitude and variable light curves make the solutions not well determined in spots. Therefore, the results are in fact the mean solutions for the light curve of each sector. The q-search diagrams both show flat bottom and rugged shape which means the mass ratio values may not be well constrained. Nevertheless, the photometric solutions from the two sectors are similar in parameters, though slightly different in the mass ratios. 

The parameters from {the ``s16'' solution were adopted in the RV fit, which was proved to work well. No significant third light was found in spectra (CCFs) or photometric solutions. However, a small {amount of third light is not rejected yet since photometrically the orbital inclination is quite low, and spectroscopically the SNR is not high which makes the CCFs a little bit of low quality. Nevertheless, the W-subtype configuration is undoubtedly confirmed. The masses of the components determined from the joint analysis are quite small compared with other targets, which makes it an interesting object similar to BH Cas \citep{Zola01}.

\section{Summary and Discussion}\label{Secsum}
{The A- and W-subtype classification is a widely accepted concept in defining and describing late-type binaries, due to their distinct distribution in the mass ratio and the effective temperature (see, e.g. \citealt{Yakut05, Latkovic21b}). It has become an unavoidable topic in the study of W UMa binaries, especially in tracing their evolutionary routes and status (see, e.g. \citealt{LiL08, Yildiz13}). However, the detailed model of W UMa binaries is under debate (e.g. \citealt{Stepien09}), even for the mechanism of the two subtypes (see \citealt{ZhangX20} and \citealt{SongH20}). The K-type CBs are essential for the study of the A/W subtype problem since they have the shortest orbital periods among W UMa binaries. The W-subtype phenomenon of K-type CBs is so special among late-type binaries that it might provide key clues for understanding the A- and W-subtype nature and thus the structure and evolution of W UMa binaries.

\subsection{W-subtype contact binary systems}
The RVs were determined {for the first time for six selected early K-type targets using spectroscopic data mainly from APOGEE and LAMOST. The RV data were utilized to check the photometric solutions of these targets. The light curves used are mainly from TESS and are of high precision. In order to make {a comparative study, the photometric solutions were initially derived independently and then checked in the fitting of RVs. It is proved that all the selected targets are W-subtype shallow CBs, which were selected unconcerned of subtypes. It is hardly a coincidence. We tried to collect all K-type CBs (period $>$ 0.215 d) with both photometric and spectroscopic measurements. Their parameters are listed in Table \ref{tabcollect}, together with our newly determined ones. Altogether are 24 objects, which are arranged into three parts (by the horizontal lines) in this table. Except {for the middle part which contains the six newly confirmed W-subtype systems from this paper, others are briefly described as follows. 

In the first (upper) part, all targets are W-subtype systems. There are three binaries -- i Boo, BH Cas and AH Vir to be mentioned. For i Boo, it is a special object which got an A-subtype solution first but was later confirmed by \cite{Lu01} to be {of W-subtype. This target was found in a triple system \citep{Hill89a}. At a small distance (parallax $\sim$ 80 mas) to our Sun, it is very bright. And the tertiary component becomes a visual companion at a separation about 1.7$^{\prime\prime}$. This companion (i Boo A) is, however, brighter than the binary member (i Boo B) and is hard to be excluded in observations, and thus may affect the photometric solutions. For BH Cas and AH Vir, their orbital periods are much longer than other K-type CBs listed. For BH Cas, the spectral type was speculated to be K4 and F8 \citep{Metcalfe99,Zola01} successively, and finally determined to be K3V+G8V by \cite{LiuJ19}. The absolute parameters of BH Cas were found pretty small compared with that of other binaries, while for AH Vir, the total mass is much larger. Nevertheless, all systems (in this part) are confirmed W-subtype CBs. 

The lower part contains four exceptions. Among them, OT Cnc (cross id. GSC 1387-0475) and J0930B belong to a special situation that the temperature difference between the two components \text{is} quite small, close to their errors. Therefore, the A/W classifications become meaningless for them according to the definition. They are labeled ``A/W'' in the table to refer to those with almost identical temperatures for the two binary components. VZ Psc may belong to another situation. It was considered a member of a small subgroup of CBs referred to as the B-subtype \citep{Lucy79, Maceroni90} which show {a large surface temperature difference between two component stars. Though with almost the same period and similar shape of light curves, this object was finally found more favorable of a near contact configuration with a fill-out factor of $-5\%\pm9\%$ \citep{Hrivnak95}. The last ``exception'' listed is V345 Gem. Although it has a period {of only $\sim$ 0.27 days, the effective temperature of the primary is over 6000 K, which is obviously not K-type. Therefore, it is not {an exception for the W-subtype phenomenon of K-type CBs.

From the above analysis, we found that there is not even one sample in Table \ref{tabcollect} that really belongs to {the A-subtype. Therefore, it is concluded that almost all of the K-type CBs might be in W-subtype configuration except a few systems which have identical temperatures for their components and thus are not necessarily to be classified as W- or A-subtype according to the original definition. The difference between the primary and the secondary components ($T_p - T_s$) is usually a few {hundred Kelvins (negative values). A few samples (labeled as ``A/W'') have positive temperature difference which is rather small (usually dozens of Kelvins). The word ``almost'' is used because we {lack sufficient K-type CBs. Although samples from Table \ref{tabcollect} cover many situations, including a large period range, low orbital inclinations, and relatively small mass ratios, many cases are not covered yet, such as {extremely low mass ratios ( $M_s / M_p < 0.15$), very low inclinations ($i < 40$), ultra-short periods below the cut-off ($\sim 0.22$ d), systems with evolved components (the example is BH Cas \citep{Zola01}). More observations are necessary in the future, especially the spectroscopic observations for RV determination, though recent studies from photometric {data alone {have already found many late-type CBs (with periods close to the cut-off) overwhelmingly in W-subtype \citep{LiK19,LiK20,Latkovic21a}.

\subsection{Additional comments}
Three of our targets are possible triple systems, among which V384 Ser and RV CVn are quite certain. For V384 Ser, the spectra (CCFs) proved the existence of a tertiary component. The contribution of the third light matches the photometric solutions quite well and it is also  consistent with the third body calculation deduced by the O-C diagram of the orbital period study \citep{Michaels19}. Therefore, this target is very likely a hierarchical triple system. For RV CVn, a close companion (2MASS J13401740+2818208) is only $\sim~10^{\prime\prime}$ away and about one percent as bright as RV CVn. With nearly identical distance (parallax = $2.37\pm0.16$ mas according to Gaia DR2 \citep{Gaia18,Luri18}) and very similar {proper motions ((-6.32,-7.42) mas/yr for RV CVn and (-6.31,-7.01) mas/yr for the companion), this visual companion is likely to be a physical one. V1038 Her is another possible triple system, which is supported by both photometric solutions and the spectroscopic study, which are in good agreement. For the other three targets, no distinctive third lights are detected in their photometric solutions nor in the spectroscopic study. So, the detecting rate of a third light in our samples is similar to that of other CBs collected in Table \ref{tabcollect} (9 in 18 samples).

Finally, the physical parameters of the targets listed in Table \ref{tababs}, however, are still roughly determined values based on the following factors. Firstly, the spectroscopic observations for these targets are not numerous enough, and the phase coverage is not ideal either, which is the reason that the mass ratio cannot be determined with spectra alone and should be determined photometrically in the beginning. The typical errors of the mass ratios of our targets are thereby relatively large, about 0.03--0.05 ($1/q$) (this won't change the W/A subtype results). Secondly, the orbital inclinations are from the photometric solutions, which are correlated with the lengths of the semi-major axis and thus affect the absolute parameters. Nevertheless, the orbital inclinations are usually well constrained by the light curves. Therefore, the uncertainties in inclinations only moderately increase the errors of physical parameters. {Thirdly, there might be various sources of systematic errors, including (but not limited to) the smear effect of LLCs, the long exposure time of the spectroscopic data, and the single broad-band photometry. In all, the real uncertainties should be larger than the standard errors listed in Table \ref{tababs} which {are only based on the fitting. Since the uncertainties of absolute parameters are huge, they should be of limited use. For the last column, ``$L_{G2}$'' values are calculated using the magnitudes in V from database and the distance based on Gaia DR2. Taking account of various errors involved, the total luminosities are generally in agreement with luminosities based on Gaia DR2.

\acknowledgments

The data presented in this paper were obtained mainly from TESS mission and SDSS-IV. Funding for the TESS mission is provided by NASA Science Mission directorate. We acknowledge the TESS team for its support of this work. This work also includes data from AAVSO. We acknowledge with thanks the variable star observations from the AAVSO International Database contributed by observers worldwide and used in this research.

Funding for the Sloan Digital Sky Survey IV has been provided by the Alfred P. Sloan Foundation, the U.S. Department of Energy Office of Science, and the Participating Institutions. SDSS acknowledges support and resources from the Center for High-Performance Computing at the University of Utah. The SDSS web site is www.sdss.org.

SDSS-IV is managed by the Astrophysical Research Consortium for the 
Participating Institutions of the SDSS Collaboration including the 
Brazilian Participation Group, the Carnegie Institution for Science, 
Carnegie Mellon University, the Chilean Participation Group, the French Participation Group, Harvard-Smithsonian Center for Astrophysics, 
Instituto de Astrof\'isica de Canarias, The Johns Hopkins University, 
Kavli Institute for the Physics and Mathematics of the Universe (IPMU) / 
University of Tokyo, the Korean Participation Group, Lawrence Berkeley National Laboratory, 
Leibniz Institut f\"ur Astrophysik Potsdam (AIP),  
Max-Planck-Institut f\"ur Astronomie (MPIA Heidelberg), 
Max-Planck-Institut f\"ur Astrophysik (MPA Garching), 
Max-Planck-Institut f\"ur Extraterrestrische Physik (MPE), 
National Astronomical Observatories of China, New Mexico State University, 
New York University, University of Notre Dame, 
Observat\'ario Nacional / MCTI, The Ohio State University, 
Pennsylvania State University, Shanghai Astronomical Observatory, 
United Kingdom Participation Group,
Universidad Nacional Aut\'onoma de M\'exico, University of Arizona, 
University of Colorado Boulder, University of Oxford, University of Portsmouth, 
University of Utah, University of Virginia, University of Washington, University of Wisconsin, 
Vanderbilt University, and Yale University.

A few spectroscopic data were provided by Guoshoujing Telescope (LAMOST), which is a National Major Scientific Project built by the Chinese Academy of Sciences. Funding for the project has been provided by the National Development and Reform Commission. LAMOST is operated and managed by the National Astronomical Observatories, Chinese Academy of Sciences.

This research has made use of the International Variable Star Index (VSX) database, operated at AAVSO, Cambridge, Massachusetts, USA. This research has also made use of the SIMBAD and Vizier databases. This work also presents a few results from the European Space Agency (ESA) space mission Gaia. Gaia data are being processed by the Gaia Data Processing and Analysis Consortium (DPAC). Funding for the DPAC is provided by national institutions, in particular the institutions participating in the Gaia MultiLateral Agreement (MLA).

This publication is supported by the National Natural Science Foundation of China (Nos. 11933008, 11873017, 12073069), and the Yunnan Natural Science Foundation (No. 202001AT070091). Z.-L. Yuan is supported by the Xiaoxiang Scholars Programme of Hunan Normal University. We would also thank Dr. Liang Wang at NIAOT, CAS for important help in data analysis. We are grateful to the anonymous referee for valuable advices which have improved the manuscript greatly.

\software{astropy \citep{Astropy13}
          }

\appendix

\section{The CCF profiles of targets}

The CCF profiles of targets and their multi-gaussian fittings are shown in Fig. \ref{Figccf1} and \ref{Figccf2}.

\section{Newly determined minimum light times from TESS data}
New minimum light times were determined for the light curves extracted from TESS. Because the data are sparsely sampled for these targets (few data in one cycle), each derived time was determined using the data in a few continuous cycles and thus is an averaged result for these cycles \citep{LiuN20}. The determined times are listed in Table \ref{tabmin} for all the targets. In this table, "NA" is the number of data used for determination. The tables will be available in their entirety in machine-readable form.

\clearpage

\begin{figure*}
\centering
\includegraphics[width=16cm]{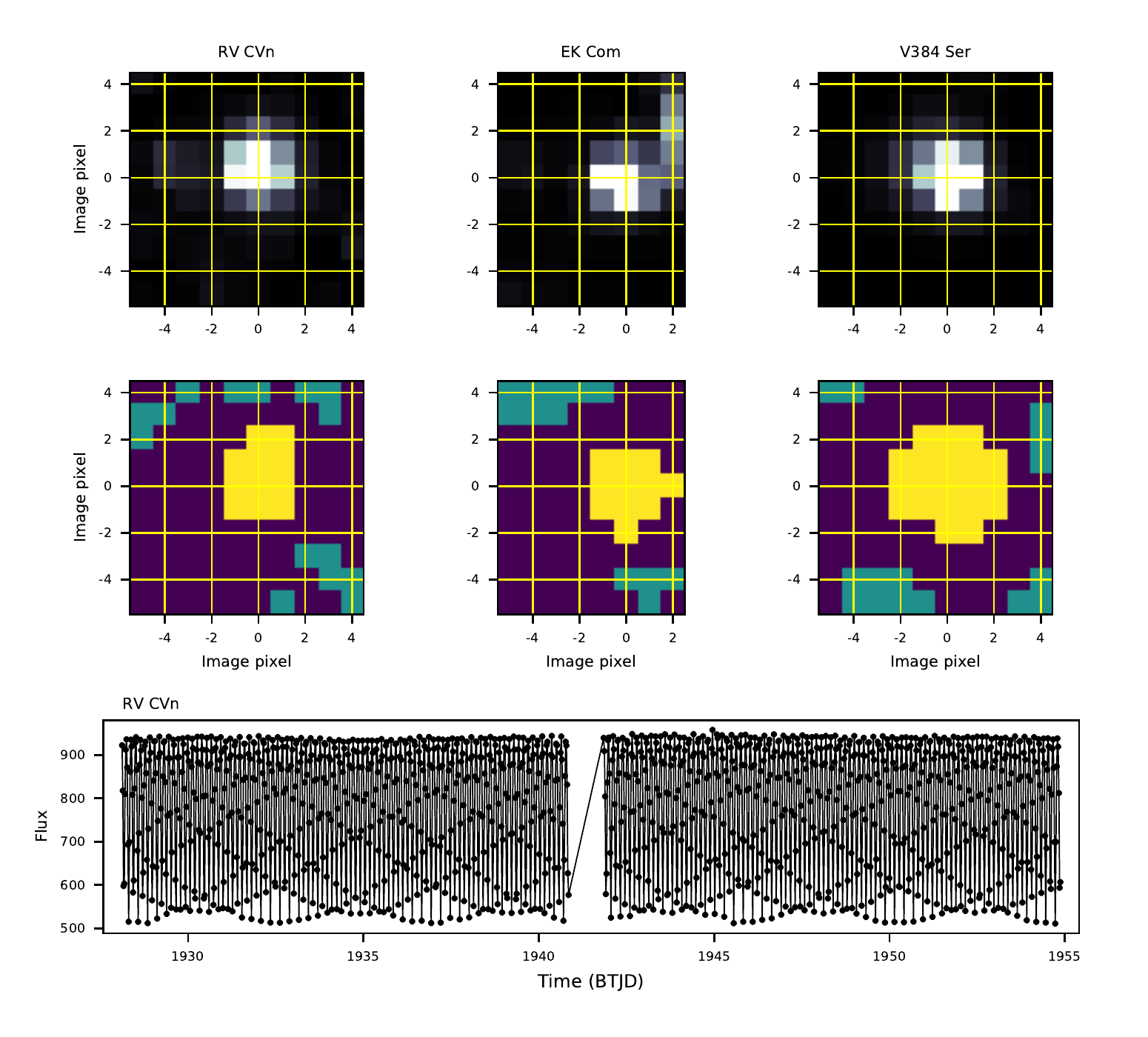}
\caption{The aperture photometry of targets. The top and middle panels show the observed images and the corresponding apertures (yellow for the aperture and blue for the background. color version online), respectively. The extracted light curve of RV CVn is shown in the lower panel. }
\label{Figaper}
\end{figure*}

\begin{figure*}
\centering
\begin{minipage}{0.75\linewidth}
\vspace{2pt}
\centerline{\includegraphics[width=\textwidth,height=5cm]{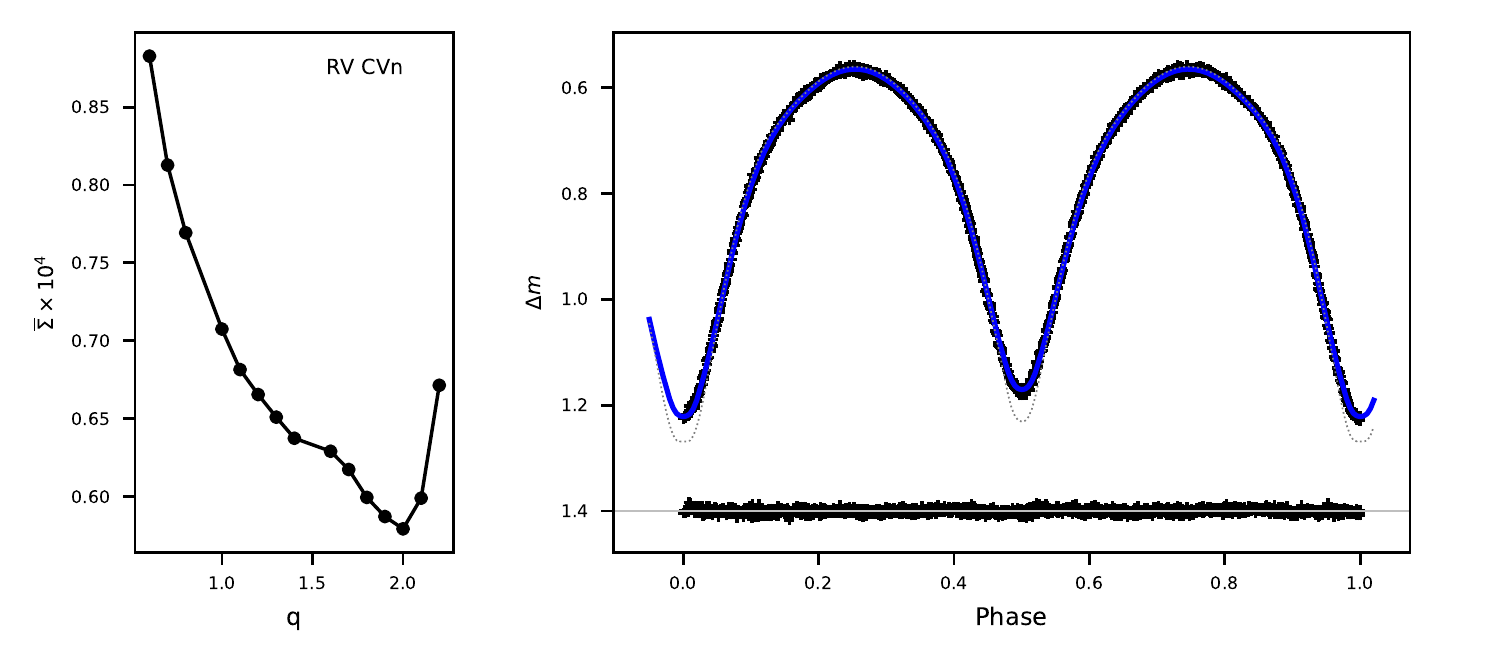}}
\vspace{2pt}
\centerline{\includegraphics[width=\textwidth,height=5cm]{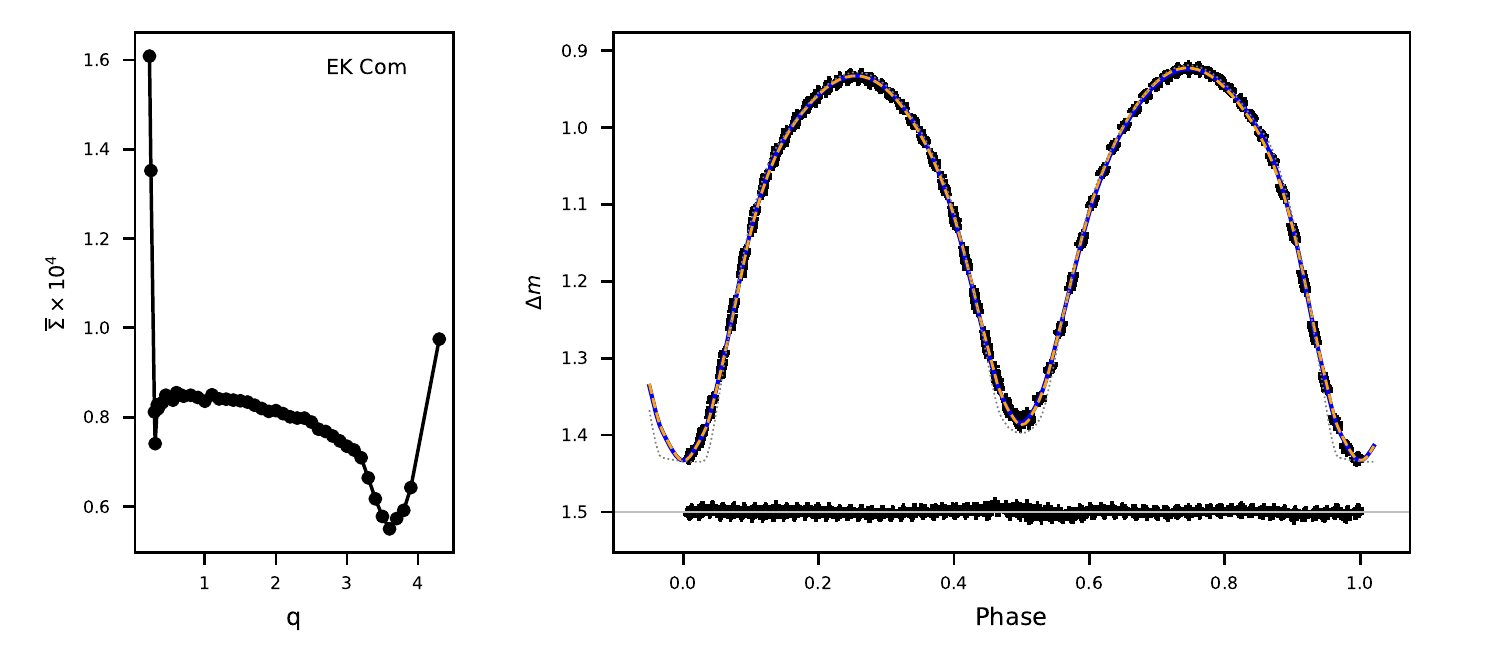}}
\vspace{2pt}
\centerline{\includegraphics[width=\textwidth,height=5cm]{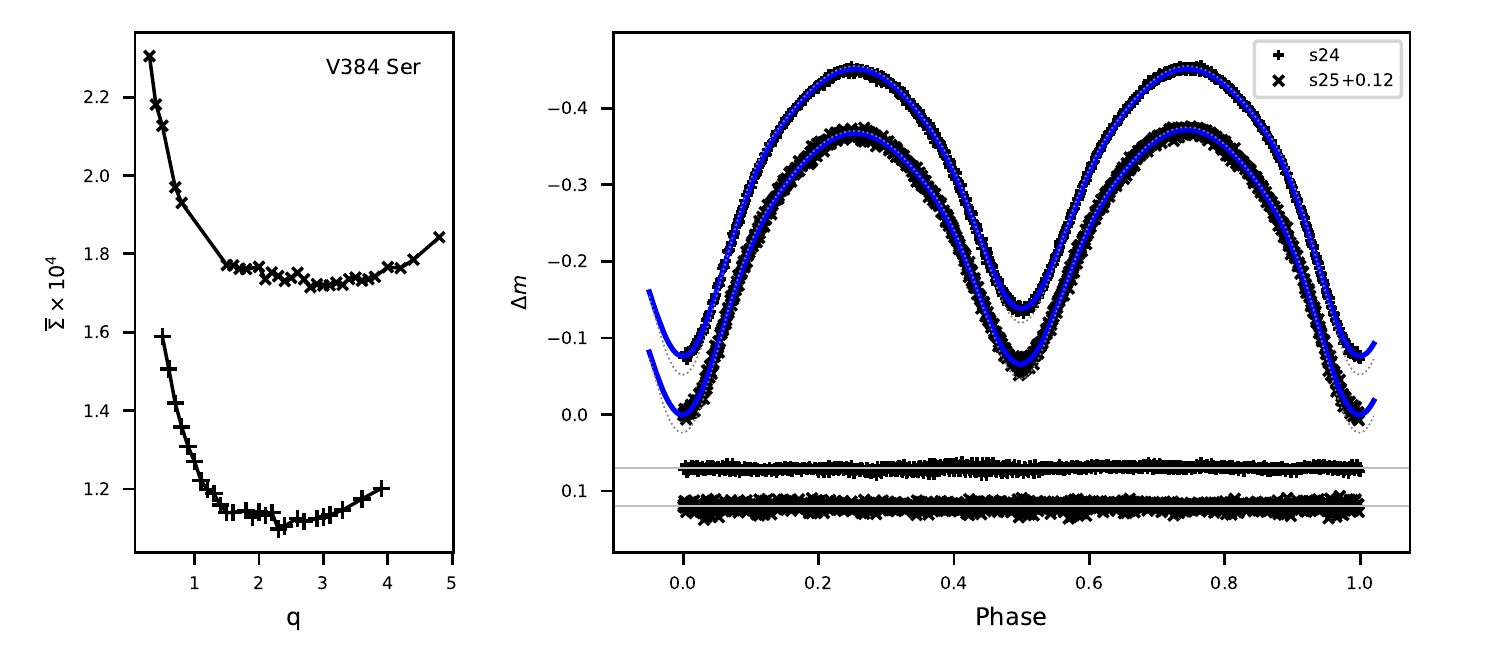}}
\end{minipage}
\caption{The $q$-$\overline{\Sigma}$ curves (left panels) and phased light curves (right panels) for RV CVn, EK Com, and V384 Ser. The target names are listed in the left panels. In the right panels, the symbols and the blue solid lines (thick) denote the observed and the synthetic (best solutions) light curves, while the grey dotted lines (thin) represent the theoretical light curves without smear effect. In the middle right panel, the calculated curve from optional results is shown by the orange dashed line. The residuals of the light curves are shown at the bottom of each panel. The label ``s'' denotes the TESS sector. ( color version online.)}\label{Figlcs1}
\end{figure*}

\begin{figure*}
\centering
\begin{minipage}{0.75\linewidth}
\vspace{2pt}
\centerline{\includegraphics[width=\textwidth,height=5cm]{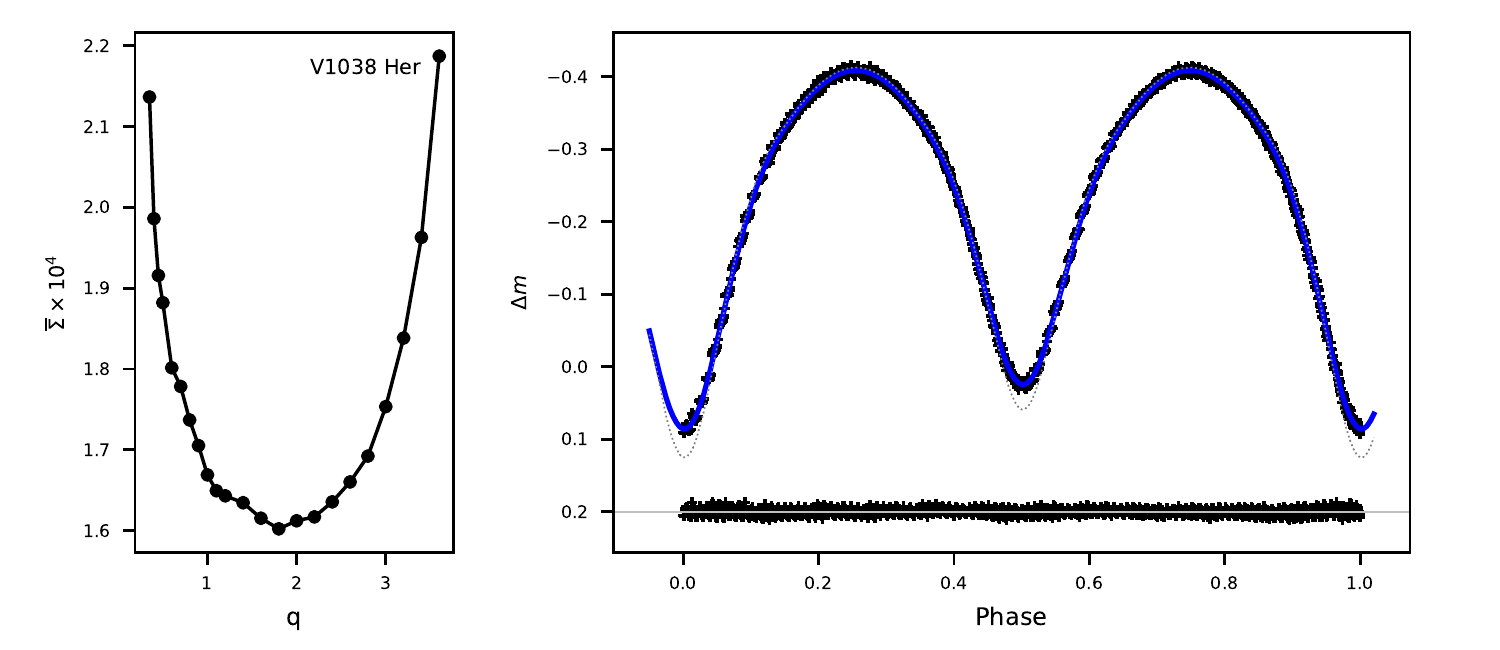}}
\vspace{2pt}
\centerline{\includegraphics[width=\textwidth,height=5cm]{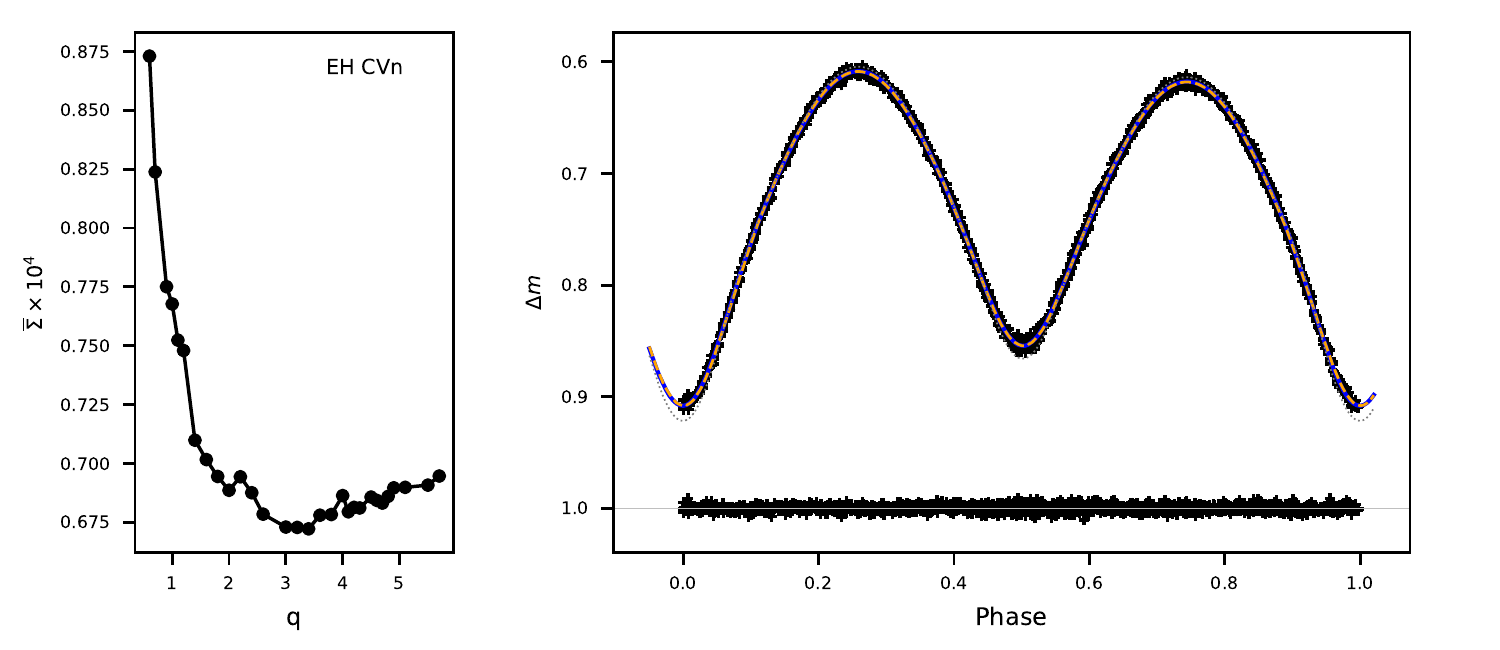}}
\vspace{2pt}
\centerline{\includegraphics[width=\textwidth,height=5cm]{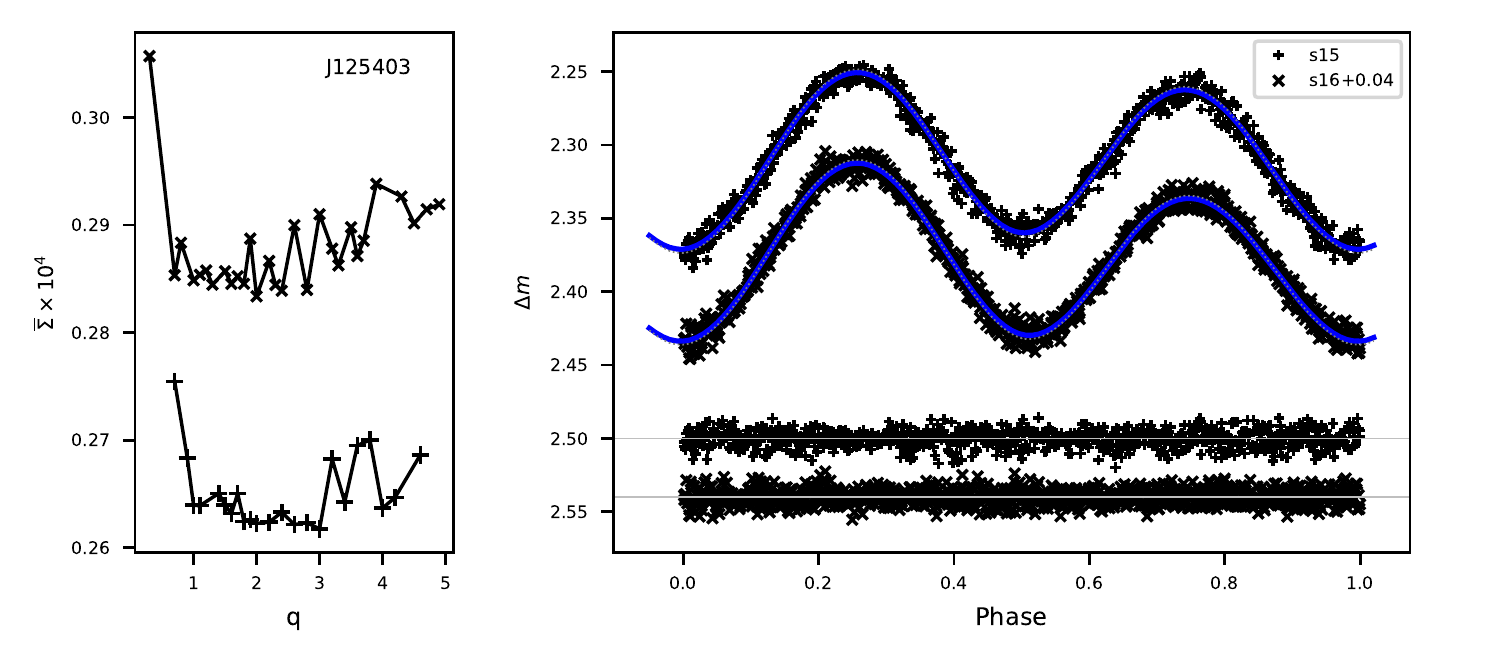}}
\end{minipage}
\caption{The same as in Figure \ref{Figlcs1}, but for V1038 Her, EH CVn, and J125403.}\label{Figlcs2}
\end{figure*}

\begin{figure*}
\centering
\includegraphics[scale=0.72]{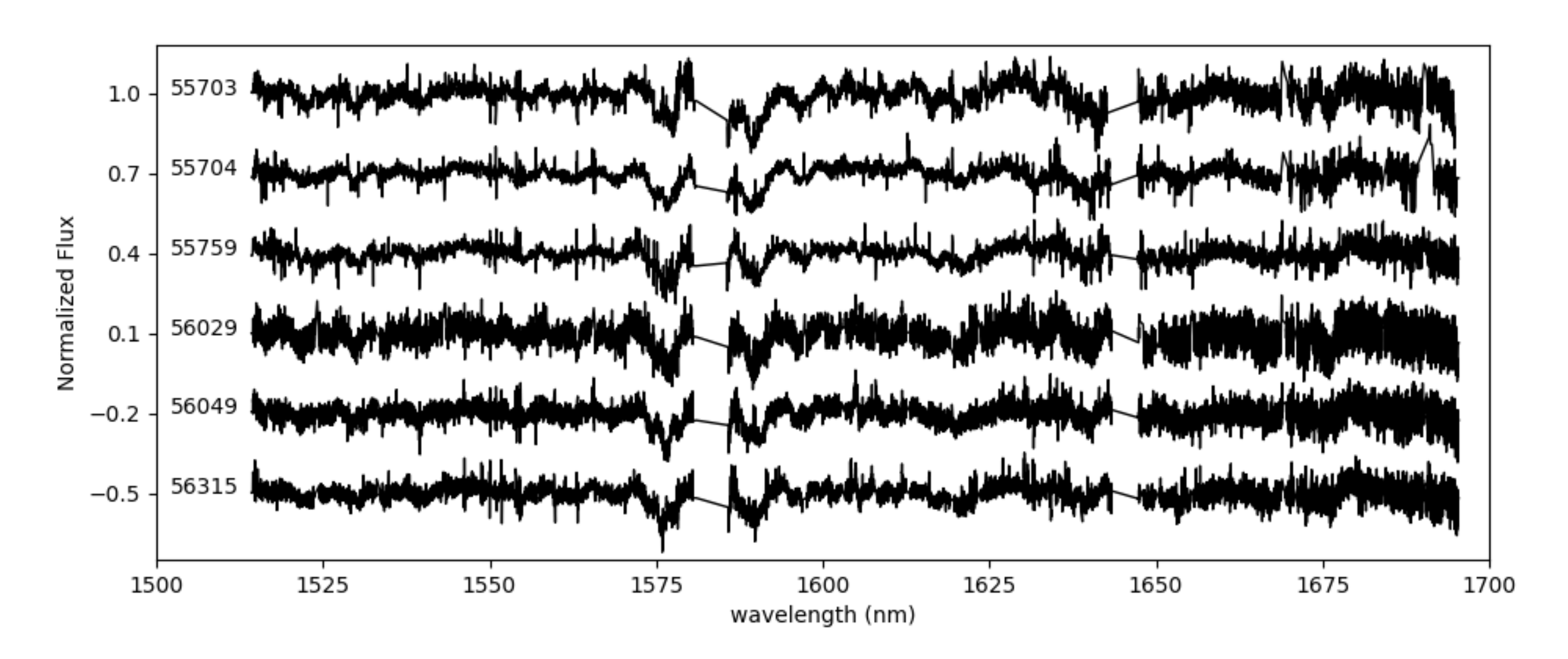}
\caption{The normalized spectra of RV CVn from APOGEE-1. The spectra are vertically shifted for clarity. MJD of each spectrum is marked on the left side.}
\label{Figspn}
\end{figure*}

\begin{figure*}
\centering
\subfigure{
\begin{minipage}{0.45\linewidth}
\vspace{2pt}
\centerline{\includegraphics[width=\textwidth]{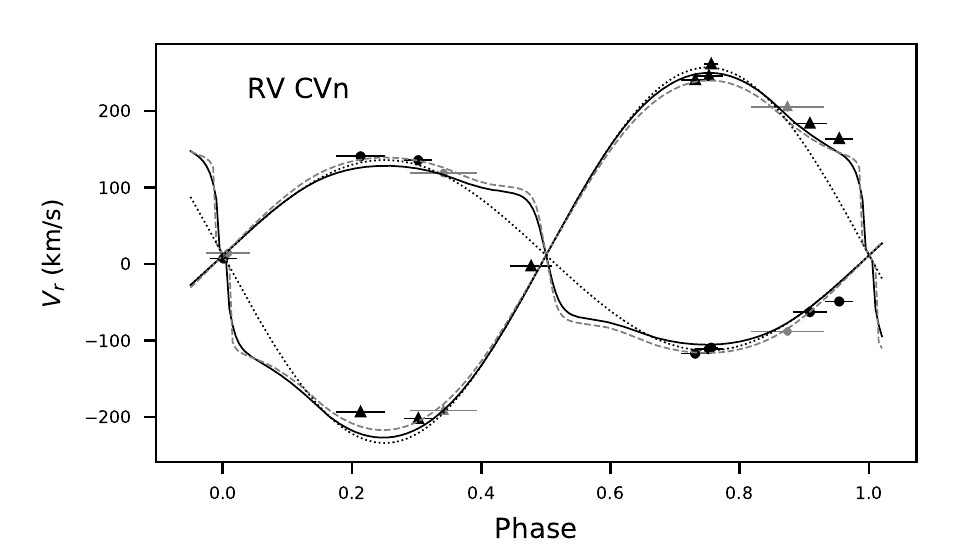}}
\vspace{2pt}
\centerline{\includegraphics[width=\textwidth]{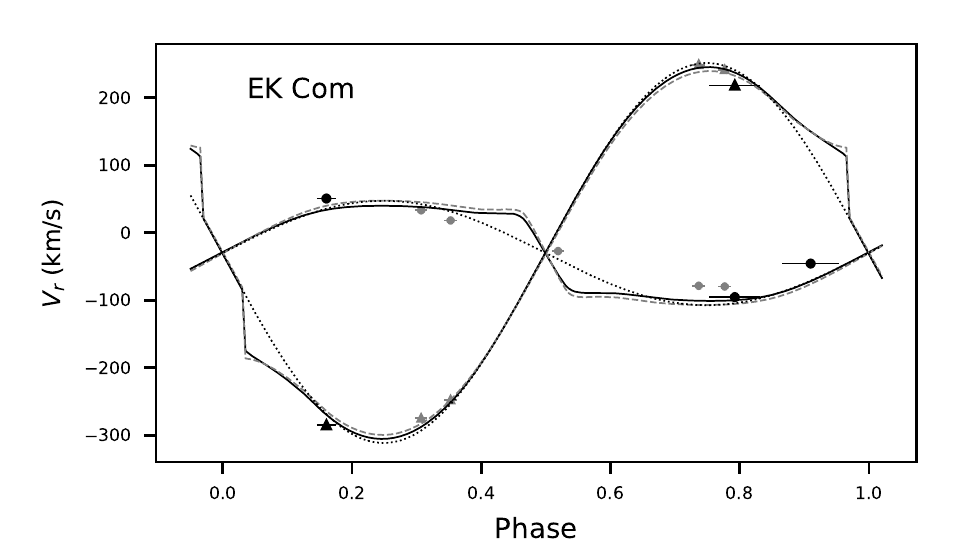}}
\vspace{2pt}
\centerline{\includegraphics[width=\textwidth]{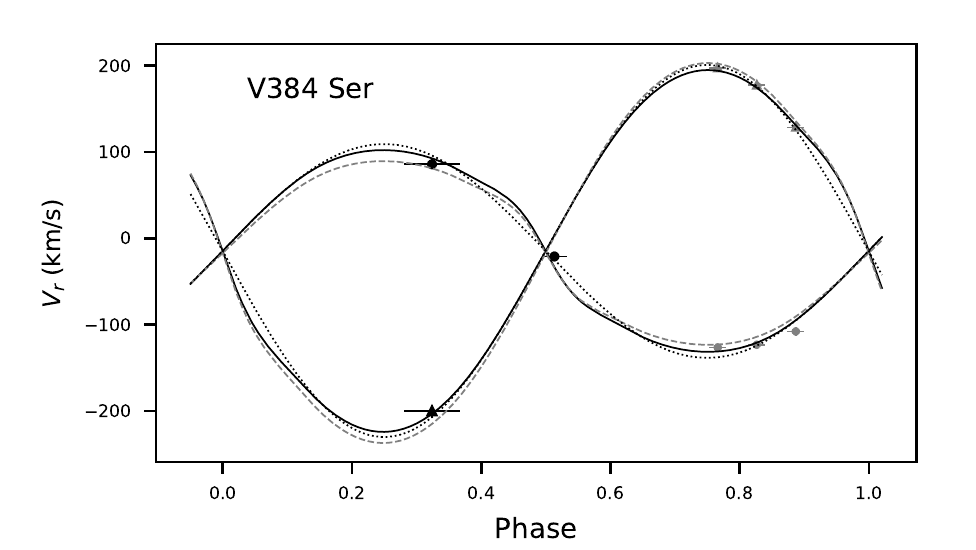}}
\end{minipage}
}
\subfigure{
\begin{minipage}{0.45\linewidth}
\vspace{2pt}
\centerline{\includegraphics[width=\textwidth]{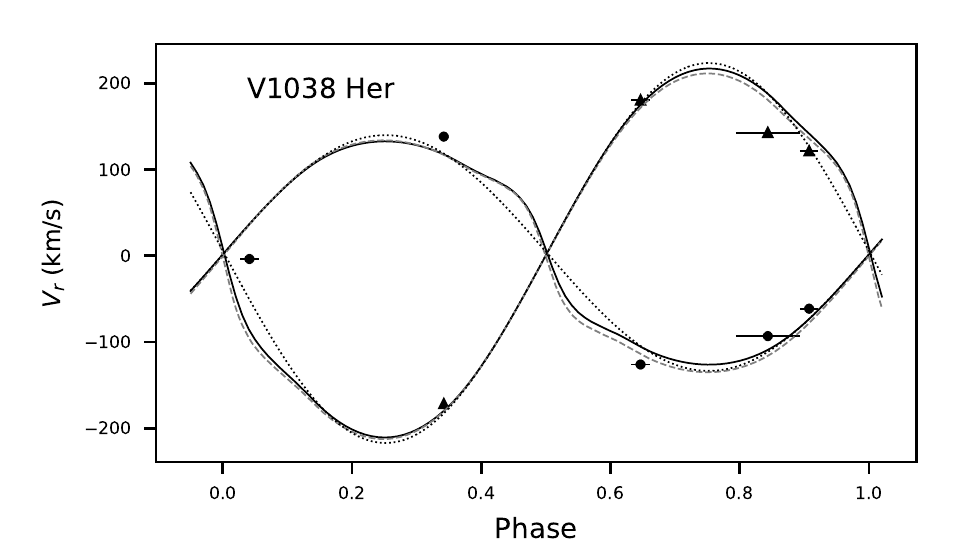}}
\vspace{2pt}
\centerline{\includegraphics[width=\textwidth]{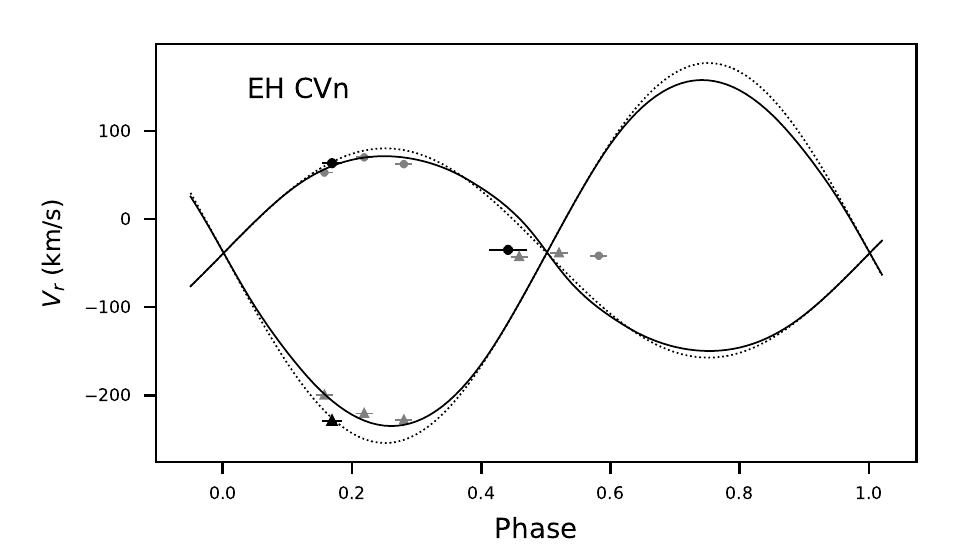}}
\vspace{2pt}
\centerline{\includegraphics[width=\textwidth]{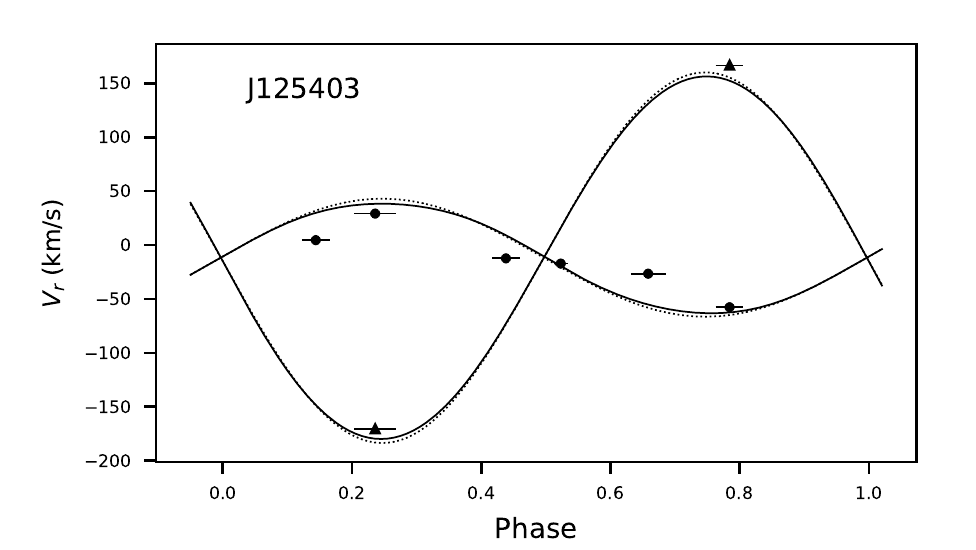}}
\end{minipage}
}
\caption{RVs of the targets and the fittings. The filled circles and triangles represent radial velocity observations of primary and secondary components, respectively. Small grey symbols denote data from different sources besides APOGEE-2: APOGEE-1 for RV CVn, TNO for EK Com, and LAMOST for V384 Ser and EH CVn. The dotted lines give the respective point-circular-orbit (sine-curve) fits while solid lines give the RV curves which are calculated from the best solutions (taking account of the proximity correction) of each object. The dashed lines in a few panels denote solutions from optional results (explained in the text) for comparison. The horizontal bars stand for half the exposure time.}
\label{Figrv}
\end{figure*}

\begin{figure}
\centering
\includegraphics[width=9cm]{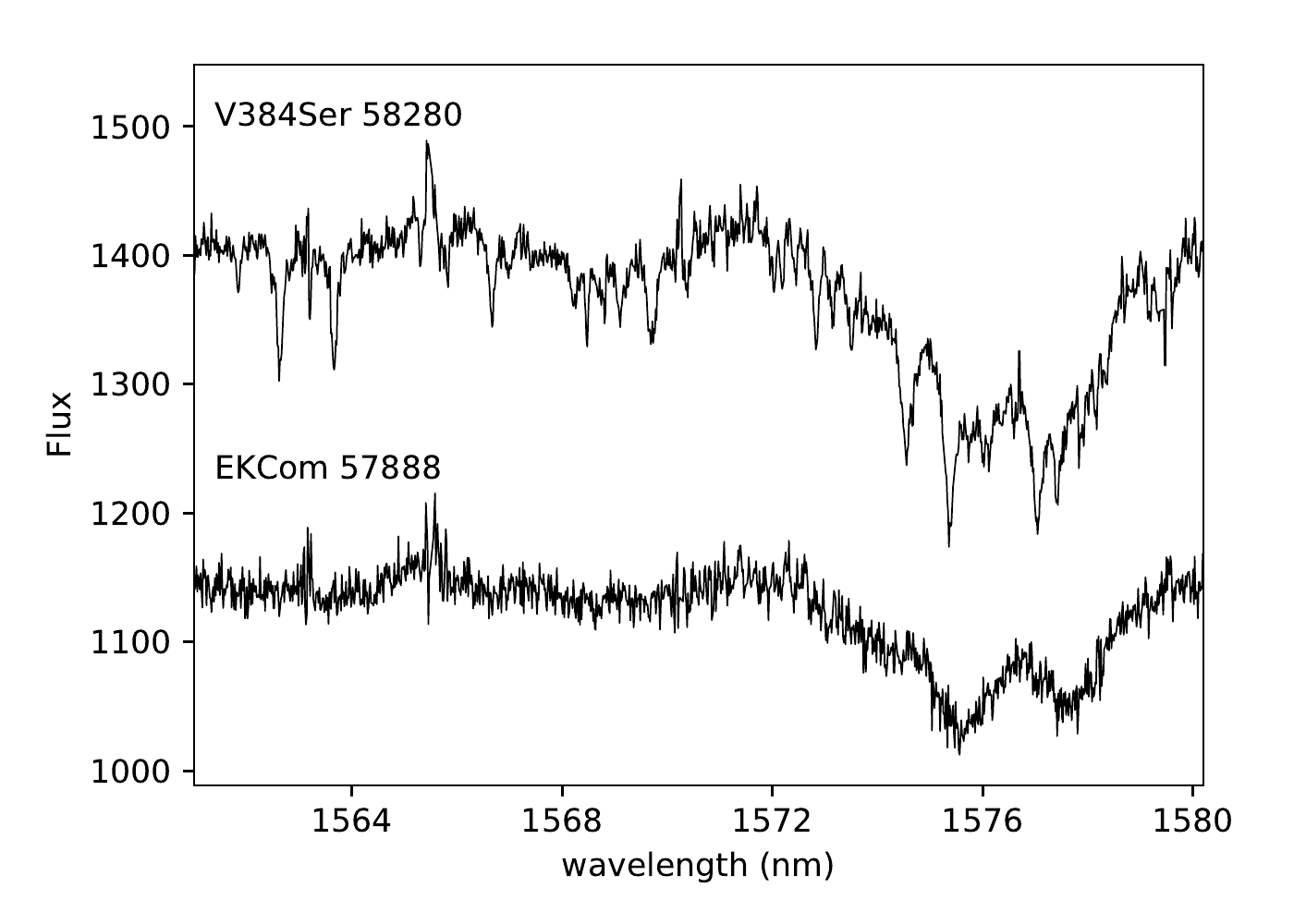}
\caption{The spectra of V384 Ser and EK Com (a zoomed-in view) from APOGEE. The lower spectrum is vertically shifted. MJDs of the spectra are listed on the left.}
\label{Figspa}
\end{figure}

\begin{figure}
\centering
\includegraphics[width=9cm]{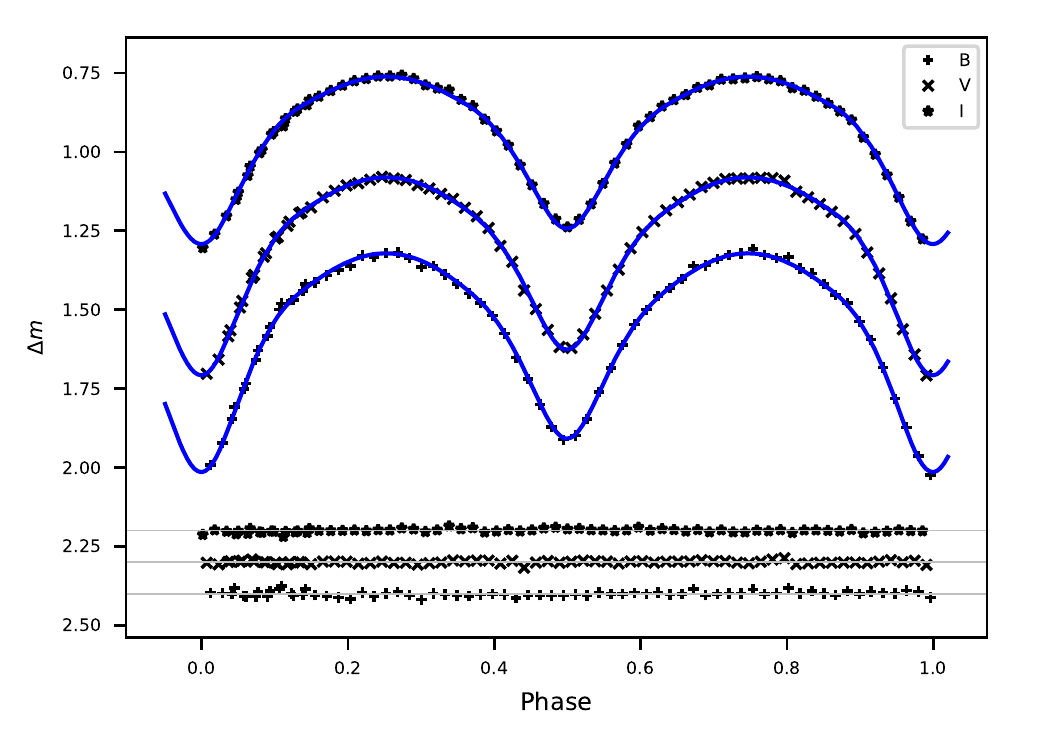}
\caption{The multi-band light curves of V1038 Her and the theoretical fittings.}
\label{figlcs_V1038}
\end{figure}

\begin{figure*}
\centering
\subfigure{
\begin{minipage}{0.46\linewidth}
\vspace{2pt}
\centerline{\includegraphics[width=\textwidth]{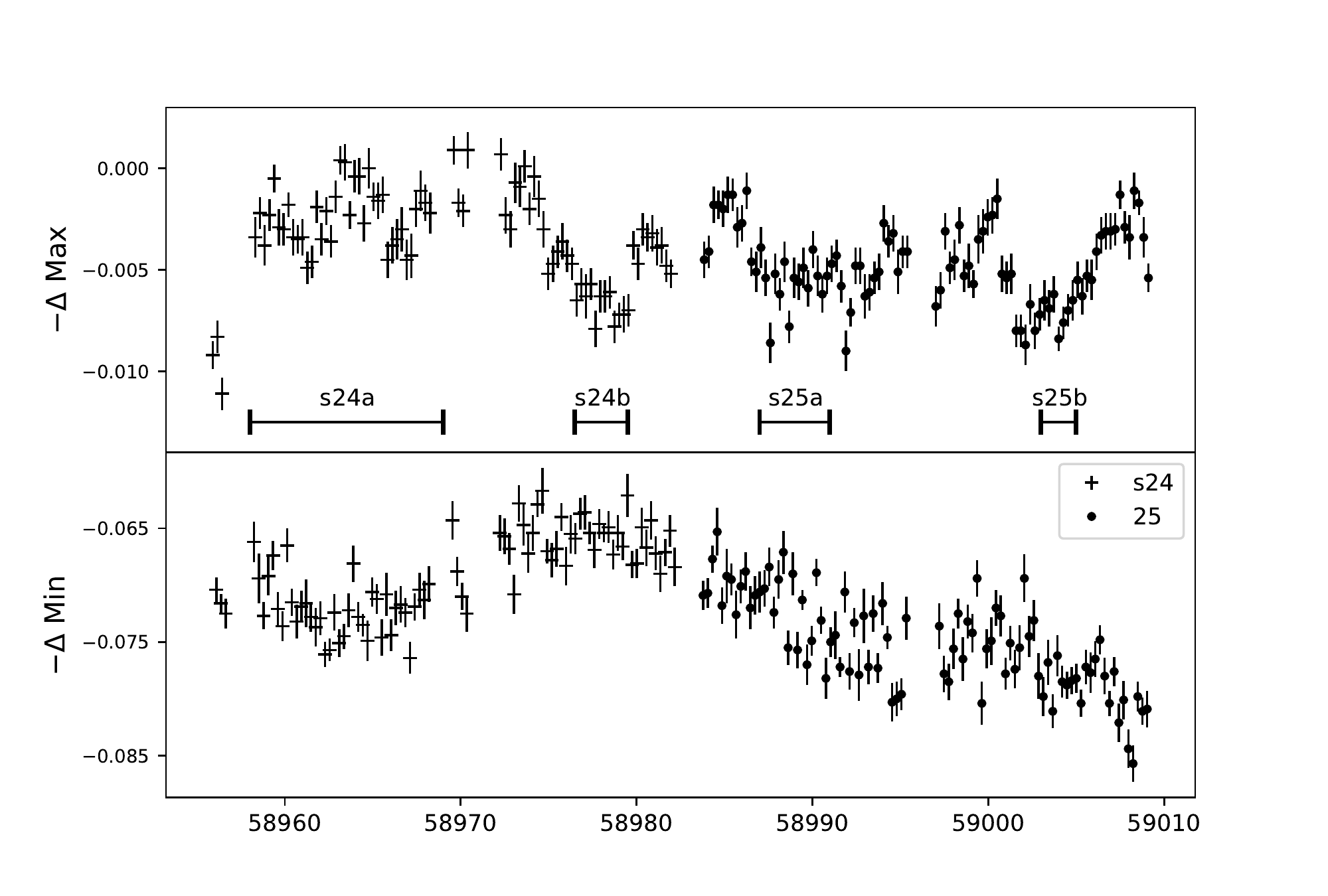}}
\vspace{2pt}
\centerline{\includegraphics[width=\textwidth]{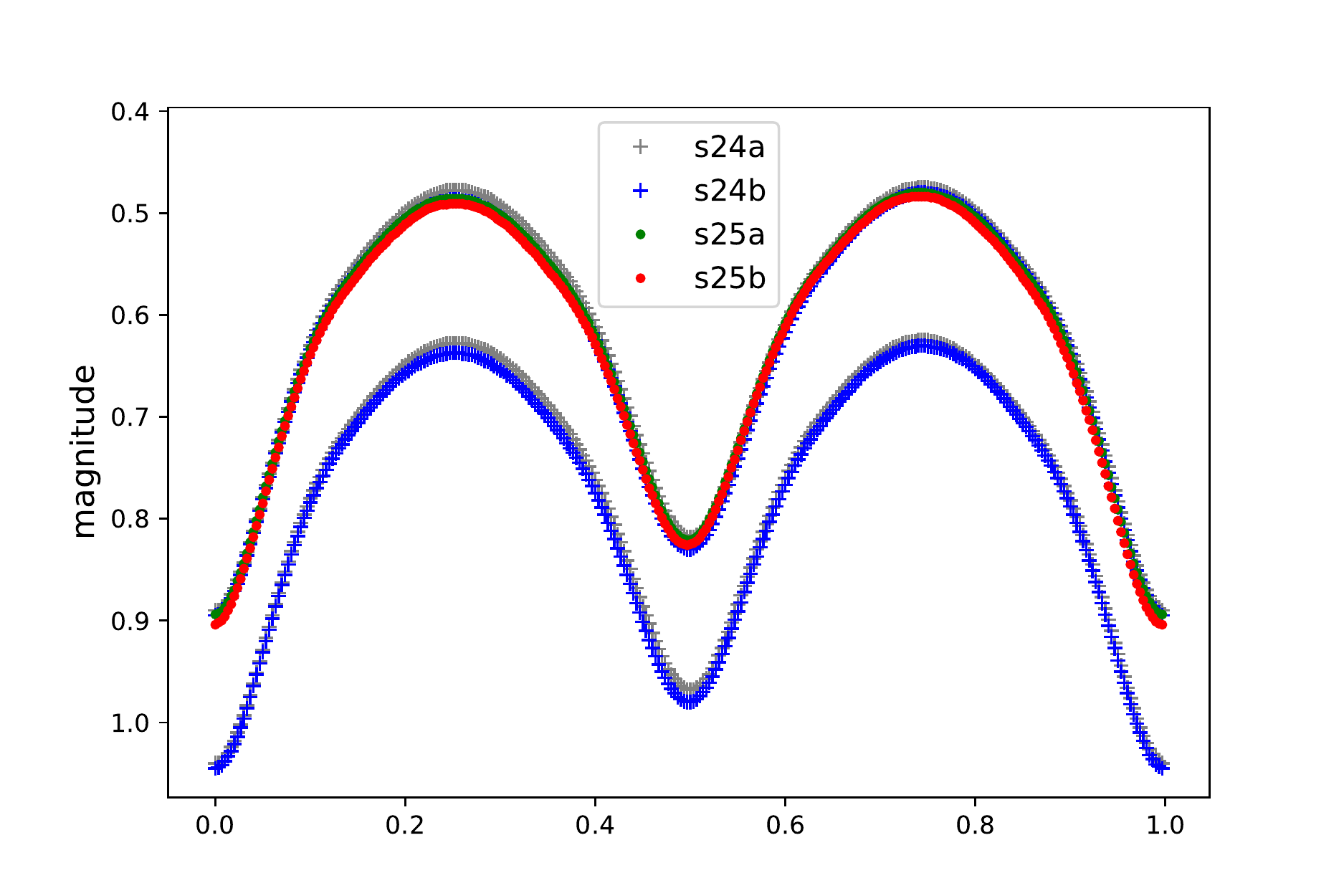}}
\end{minipage}
}
\subfigure{
\begin{minipage}{0.46\linewidth}
\vspace{2pt}
\centerline{\includegraphics[width=\textwidth]{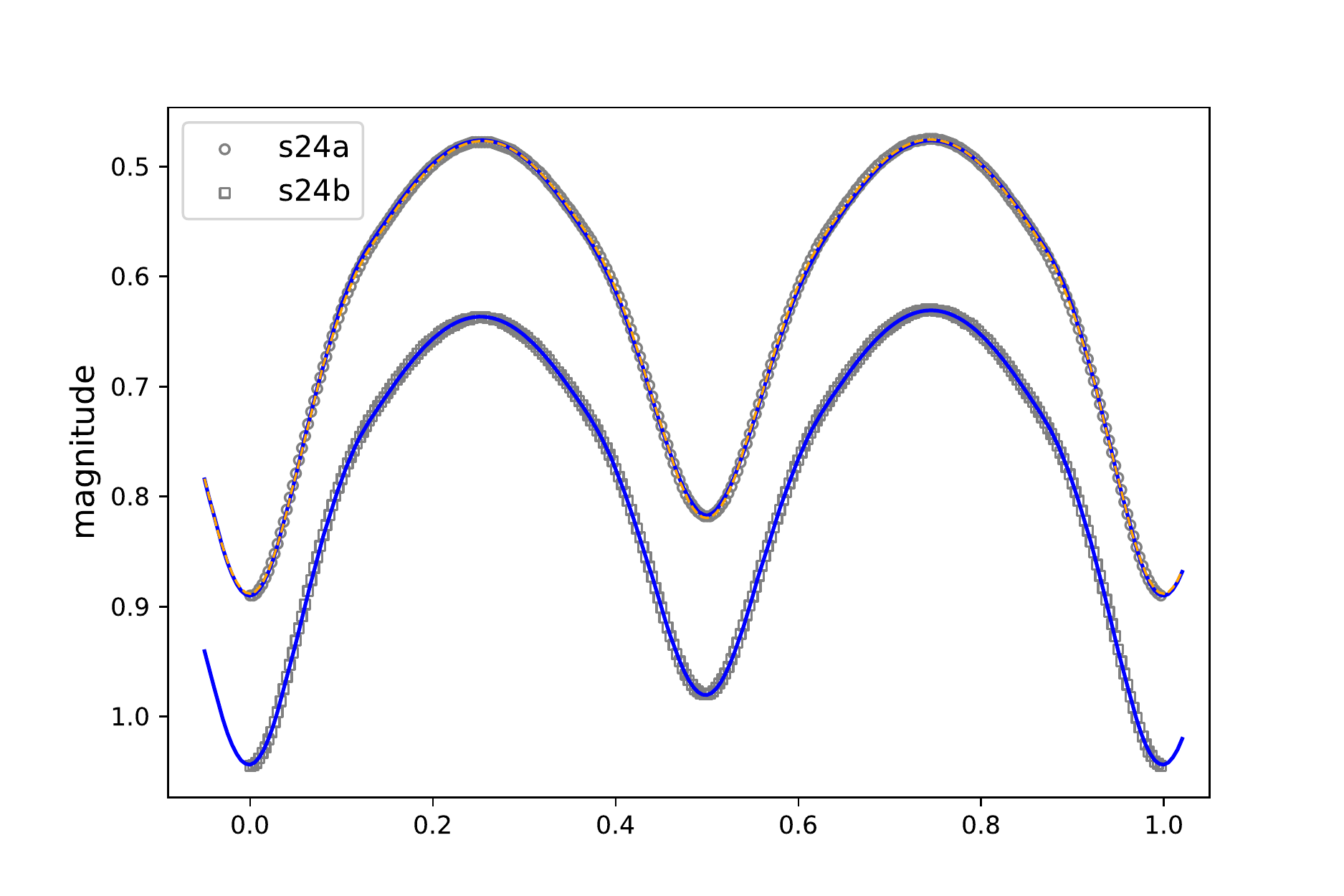}}
\vspace{4pt}
\centerline{\includegraphics[width=\textwidth]{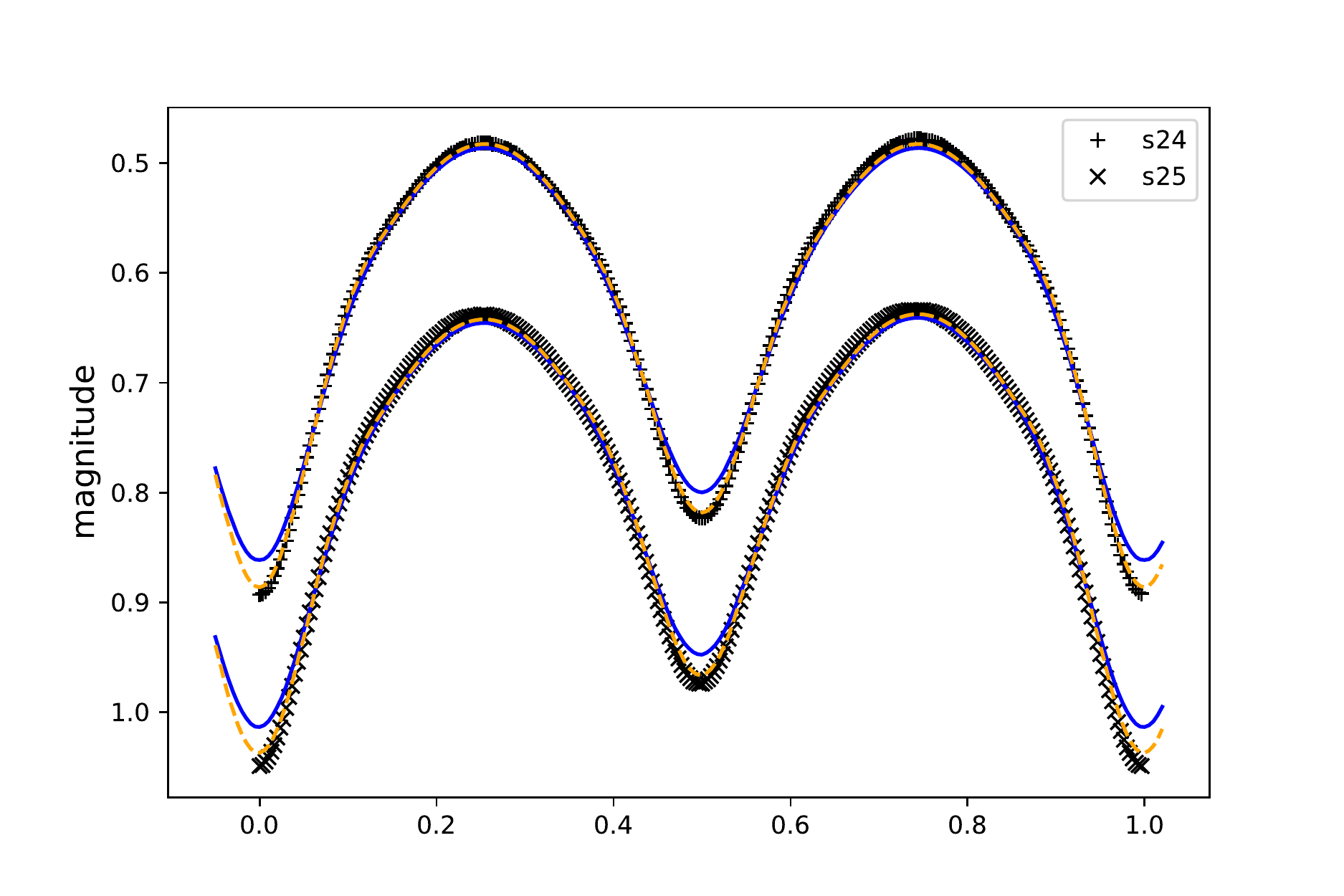}}
\end{minipage}
}
\caption{The analysis of TESS short cadence light curves of V384 Ser. Upper left: the variation of the light curves indicated from $-\Delta Max$  and $-\Delta Min$. Lower left: comparison of light curves of different segments. Upper right: the observed and calculated light curves. Lower right: the short cadence light curves and the fittings from the results of LLCs. The blue solid lines represent theoretical light curves while the orange dashed lines denote calculated light curves without smear effect. Except for the upper left panel, all light curves shown are rebinned from the original data. The light curves are vertically shifted for clarity.}
\label{Figlcsc}
\end{figure*}

\begin{figure*}
\centering
\includegraphics[]{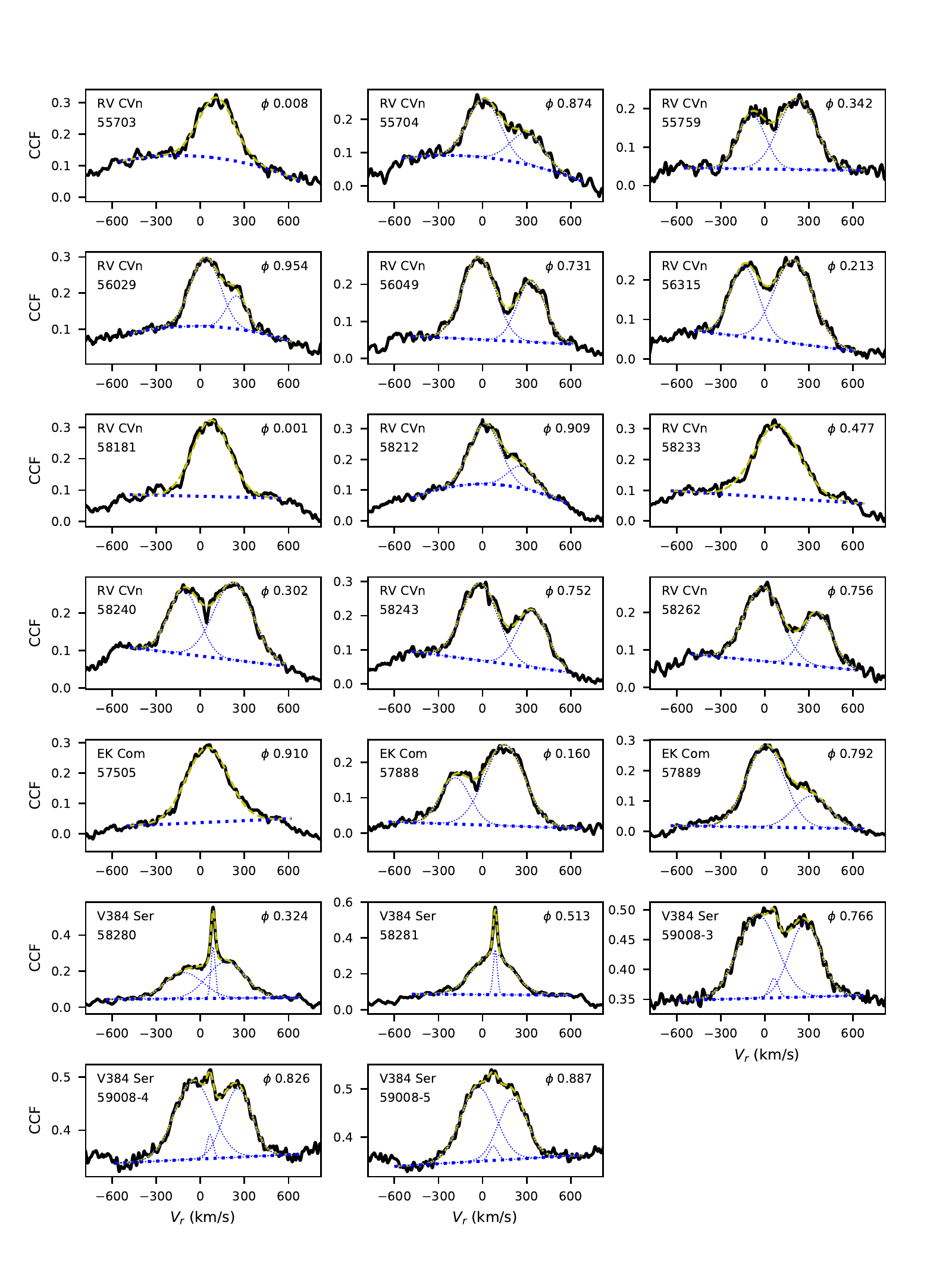}
\caption{The CCF profiles of targets and the multi-gaussian fittings for RV CVn, EK Com, and V384 Ser. The multi-gaussian components are shown in blue dotted lines (thin), while the linear or quadratic components are shown in thick dotted lines, and the final profiles fitted are shown in yellow dashed lines. The phases are shown on the upper right of each panel.} \label{Figccf1}
\end{figure*}

\begin{figure*}
\centering
\includegraphics[]{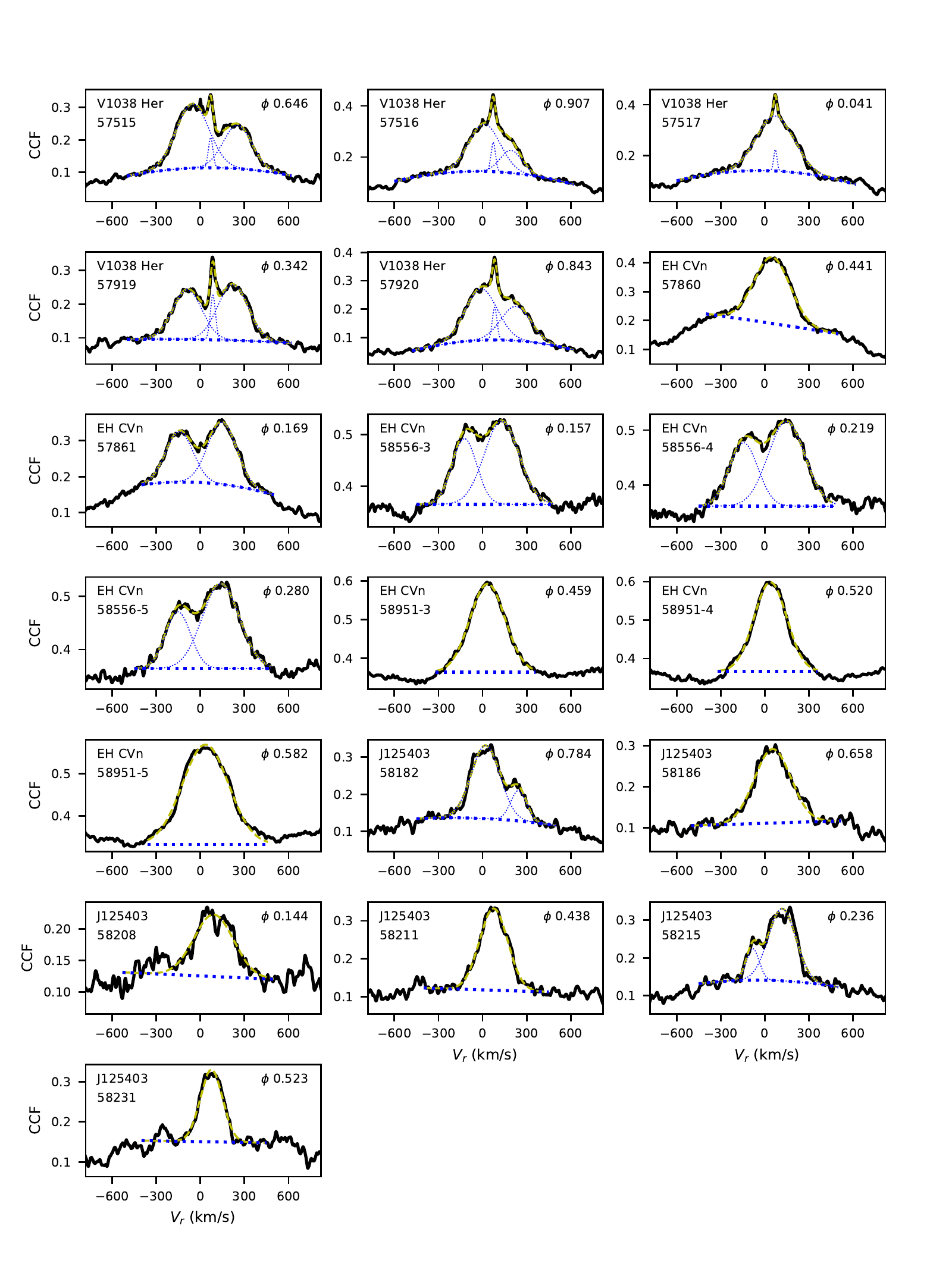}
\caption{The same as in Figure \ref{Figccf1} but for V1038 Her, EH CVn, and J125403.} \label{Figccf2}
\end{figure*}

\begin{figure*}
\centering
\includegraphics[]{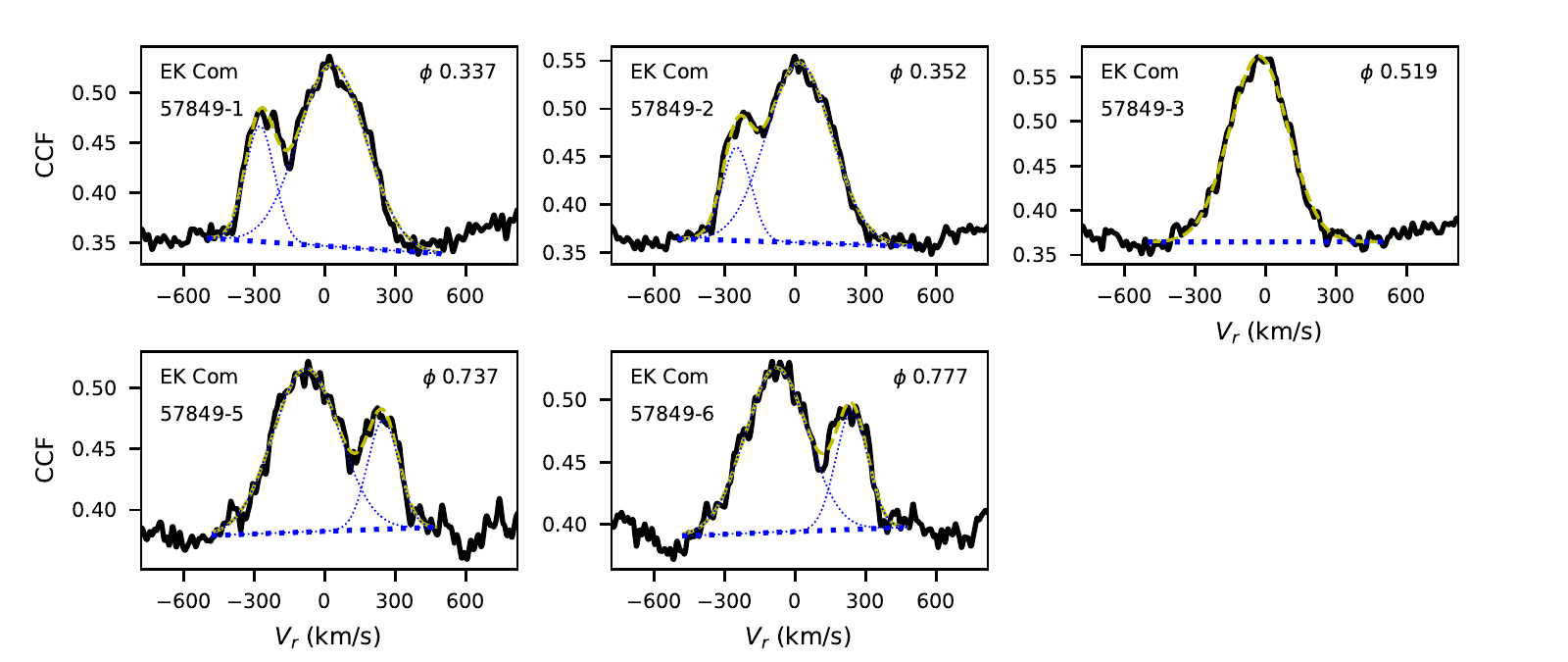}
\caption{The same as in Figure \ref{Figccf1} but for EK Com from the observations at TNO.} \label{Figccf3}
\end{figure*}

\clearpage

\begin{table*}[htbp]
\caption{Basic information of the variables from databases.}
\label{tabinfo}
\scriptsize
\centering          
\begin{tabular}{lccllcccccc}
\hline\hline  
Name      & $\alpha_{2000}$ &  $\delta_{2000}$ & $Vmag$ & Period (d) & $J-H$ & $Teff$ (K)$^{+}$ & Sp type$^{\ast}$ & $plx$ (mas)$^{+}$ & $plxe$ (mas)$^{+}$ 
& Sector$^{\#}$ \\\hline
RV CVn    & 13 40 18.16   & +28 18 21.5  & 13.84 &  0.269567  & 0.567 & 4853 & ---   & 2.346$\pm$0.022 & 2.408$\pm$0.019 & 23    \\ 
EK Com    & 12 51 21.44   & +27 13 47.0  & 12.166 & 0.266685  & 0.444 & 4958 & G8--G9 & 5.105$\pm$0.052 & 5.164$\pm$0.018 & 23    \\ 
V384 Ser  & 16 01 53.57   & +24 52 17.5  & 11.88  & 0.268729  & 0.476 & 4879 & K2    & 4.701$\pm$0.037 & 5.039$\pm$0.056 & 24,25 \\
V1038 Her & 16 58 19.78   & +33 40 21.6  & 11.90 &  0.2681798  & 0.539 & 4837 & K3    & 5.454$\pm$0.134 & 5.492$\pm$0.066 & 25    \\ 
EH CVn    & 13 41 13.70   & +31 47 24.3  & 13.00 &  0.2635829  & 0.522 & 5045 & K3--K5 & 3.697$\pm$0.029 & 3.734$\pm$0.014 & 23    \\ 
J125403   & 12 54 03.61   & +50 39 45.9  & 14.30 &  0.268798   & 0.430 & 4904 & ---   & 1.590$\pm$0.020 & 1.604$\pm$0.016 & 15,16 \\
\hline                  
\end{tabular}
\begin{flushleft}
\textbf{Note.} $^{+}$Retrieved from Gaia DR2 \citep{Gaia18} and EDR3 \citep{Gaia21,Lindegren21}. $^{\ast}$from LAMOST low-resolution survey (LRS). $^{\#}$TESS sector
\end{flushleft}
\end{table*}

\begin{table*}[htbp]
\caption{Photometric solutions for the TESS light curves of variables (I).}\label{tabwd1}
\centering          
\begin{tabular}{lll|lll|lll}
\hline\hline  

Parameters              & \multicolumn{2}{c|}{RV CVn} & \multicolumn{3}{c|}{V384 Ser} & \multicolumn{3}{c}{EK Com}\\
                        & qser & free & qser (s24)  & el3 (s24) & spot (s25) & qser & cool spot & hot spot \\
\hline
$q$ ($M_2/M_1$)              &    2.0     & 1.971(5)   & 2.3       & 1.993(9)   &  1.355(3)  & 3.6       & 3.260(13)    & 3.637(4)     \\ 
$\Omega_{in}$                &  5.2517    & 5.2103     & 5.6709    & 5.2423     & 4.3065     & 7.3997    & 6.9581       & 7.4476       \\ 
$\Omega_{out}$               &  4.6545    & 4.6140     & 5.0655    & 4.6453     & 3.7370     & 6.7731    & 6.3355       & 6.8205       \\ 
$T_{1}$ (K)                  &    4850$^a$&   4850$^a$ & 5033(2)   &   5043(5)  &   5010(8)  & 5240(3)   &   5279(4)    &   5354(8)    \\ 
$T_{2}$ (K)                  &    4706(1) &   4703(2)  & 4750$^a$  &   4750$^a$ &   4750$^a$ & 5000$^a$  &   5000$^a$   &   5000$^a$   \\ 
$i(^{\circ})$                &   84.05(4) & 84.16(15)  & 67.92(3)  & 69.71(10)  &  68.55(13) & 83.62(10) & 87.50(19)    & 84.22(9)     \\ 
$L_{1}/L_{total}(\%)$        &   37.85(2) &  37.97(11) & 37.42(3)  &  36.64(20) &   44.2(2)  & 27.59(3)  &  28.23(26)   &  28.93(8)    \\ 
$L_{3}/L_{total}(\%)$        &    ---     &   0.71(26) &  ---      &  10.2(4)   &    9.3(7)  & ---       &   6.5(7)     &   0.81(13)   \\ 
$\Omega_{1}=\Omega_{2}$      &  5.196(1)  & 5.152(8)   & 5.658(1)  &  5.207(11) &  4.292(6)  & 7.328(2)  &  6.848(6)    &   7.365(5)   \\ 
$f(\%)$                      &    9.4(2)  &   9.8(1.3) &  2.2(2)   &   5.9(1.8) &   2.5(1.0) & 11.4(4)   &  17.6(1.0)   &  13.1(3)     \\ 
$r_{1}$(pole)                & 0.3046(1)  & 0.3061(3)  & 0.2899(1) &  0.3031(2) & 0.3326(3)  & 0.2601(2) &  0.2702(2)   &   0.2601(1)  \\ 
$r_{1}$(side)                & 0.3187(2)  & 0.3202(4)  & 0.3025(1) &  0.3168(2) & 0.3484(4)  & 0.2714(2) &  0.2824(3)   &   0.2714(2)  \\ 
$r_{1}$(back)                & 0.3544(2)  & 0.3561(5)  & 0.3358(2) &  0.3513(3) & 0.3810(5)  & 0.3077(3) &  0.3210(4)   &   0.3083(3)  \\ 
$r_{2}$(pole)                & 0.4190(1)  & 0.4180(9)  & 0.4269(1) &  0.4170(3) & 0.3831(8)  & 0.4662(2) &  0.4611(2)   &   0.4676(3)  \\ 
$r_{2}$(side)                & 0.4460(2)  & 0.4449(11) & 0.4549(1) &  0.4435(4) & 0.4043(10) & 0.5033(2) &  0.4972(3)   &   0.5052(5)  \\ 
$r_{2}$(back)                & 0.4759(2)  & 0.4750(16) & 0.4825(2) &  0.4726(5) & 0.4348(14) & 0.5294(2) &  0.5248(3)   &   0.5314(6)  \\ 
$\theta_{s}(^{\circ})$       &    ---     & ---        &  ---      & ---        & 28.7$^{t}$ & ---       & 131.8$^{t}$  &   28.6$^{t}$ \\ 
$\psi_{s}(^{\circ})$         &    ---     & ---        &  ---      & ---        &    160( 2) & ---       &   69(15)     &  207.9(1.4)  \\
$r_{s}(^{\circ})$            &    ---     & ---        &  ---      & ---        &   13.2( 5) & ---       &   11.1(4)    &   11.4(3)    \\
$T_{s}/T_{\ast}$             &    ---     & ---        &  ---      & ---        & 0.80$^{t}$ & ---       &  0.80$^{t}$  &   1.28$^{t}$ \\ 
on Star                      &    ---     & ---        &  ---      & ---        &    2       & ---       &   2          &   2          \\
$\overline\Sigma\times 10^4$ & 0.5792     &  0.5776    &  1.0980   &  1.0918    &   1.7222   & 0.5499    &  0.3885      &   0.3312     \\
\hline
\end{tabular}
\begin{flushleft}
\textbf{Note.} qser: from q-search method. $^{(a)}$Assumed. $L_{total}=L_{1}+L_{2}+L_{3}$. $^{(t)}$Trial, the parameter
was fixed at a series of trial values in the calculation until the best value was found. ``s'' inside the parentheses in the head of columns refers to the TESS sector.
\end{flushleft}
\end{table*}

\begin{table*}[htbp]
\caption{Photometric solutions for the SLCs of V384 Ser.}\label{tabwd4}
\centering          
\begin{tabular}{lllll}
\hline\hline
Parameters              & s24a & s24a  & s24b  & s24a     \\
                        & el3  & spot  & spot  & with RV  \\
\hline
$q$ ($M_2/M_1$)              &    2.339(1)  &    1.533(1)  &    1.399(1)  &    1.740(17)  \\
$\Omega_{in}$                &   5.7246     &   4.5757     &   4.3749     &   4.8791      \\
$\Omega_{out}$               &   5.1182     &   3.9966     &   3.8028     &   4.2910      \\
$T_{1}$ (K)                  &     5050(2)  &     5004(4)  &     5002(3)  &     5062(31)  \\
$T_{2}^a$ (K)                &     4750     &     4750     &     4750     &     4750      \\
$i(^{\circ})$                &   69.90(6)   &   69.53(5)   &   69.67(4)   &    69.6(5)    \\
$L_{1}/L_{total}(\%)$        &    34.5(1)   &    41.4(1)   &    42.9(1)   &    40.2(1.0)  \\
$L_{3}/L_{total}(\%)$        &     8.0(3)   &     9.7(2)   &    10.4(2)   &     9.1(2.3)  \\
$\Omega_{1}=\Omega_{2}$      &   5.685(2)   &   4.551(1)   &   4.348(1)   &   4.848(20)   \\
$f(\%)$                      &     6.6(3)   &     4.3(2)   &     4.8(2)   &     5.4(3.4)  \\
$r_{1}$(pole)                &   0.2908(1)  &   0.3234(1)  &   0.3312(1)  &   0.3137(13)  \\
$r_{1}$(side)                &   0.3037(1)  &   0.3386(1)  &   0.3471(1)  &   0.3282(15)  \\
$r_{1}$(back)                &   0.3385(2)  &   0.3721(2)  &   0.3806(1)  &   0.3623(23)  \\
$r_{2}$(pole)                &   0.4303(2)  &   0.3945(2)  &   0.3871(2)  &   0.4055(24)  \\
$r_{2}$(side)                &   0.4592(2)  &   0.4174(2)  &   0.4091(2)  &   0.4301(30)  \\    
$r_{2}$(back)                &   0.4875(3)  &   0.4477(3)  &   0.4400(3)  &   0.4600(43)  \\    
$\theta_{s}(^{\circ})$       &   ---        &     34$^t$   &     11$^t$   &   ---         \\
$\psi_{s}(^{\circ})$         &   ---        &     172(1)   &     125(3)   &   ---         \\
$r_{s}(^{\circ})$            &   ---        &     9.8(2)   &    14.7(2)   &   ---         \\
$T_{s}/T_{\ast}$             &   ---        &   0.80$^a$   &   0.80$^a$   &   ---         \\
on Star                      &   ---        &     2        &     2        &   ---         \\
$\overline\Sigma\times 10^4$ &   0.4923     &   0.3013     &   0.2543     &   6.9101      \\
\hline 
\end{tabular}
\end{table*}

\begin{table*}[htbp]
\caption{Photometric solutions for the TESS light curves of variables (II).}\label{tabwd2}
\centering          
\begin{tabular}{lll|lll|lll}
\hline\hline  

Parameters              & \multicolumn{2}{c|}{V1038 Her} & \multicolumn{3}{c|}{EH CVn} & \multicolumn{3}{c}{J125403}\\
                        & qser & el3 & qser  & case 1 & case 2 & qser (s16) & spot (s15) & spot (s16) \\
\hline
$q$ ($M_2/M_1$)              &     1.8    &  1.612(5)  &     3.4    &   1.383(3)  &   1.816(2) &    3.0     &   2.583(21) &   3.139(21) \\
$\Omega_{in}$                &  4.9661    & 4.6916     &  7.1409    &  4.3499     &  4.9882    & 6.6163     &  6.0585     &  6.7998     \\
$\Omega_{out}$               &  4.3757    & 4.1089     &  6.5165    &  3.7787     &  4.3973    & 5.9974     &  5.4468     &  6.1788     \\
$T_{1}$ (K)                  &    4934(2) &   4942(3)  &    5222(6) &    4948(8)  &    5115(6) &   5202(28) &    5259(78) &    5204(55) \\
$T_{2}$ (K)                  &    4750    &   4750     &    4750    &    4750     &    4750    &   4750     &    4750     &    4750     \\
$i(^{\circ})$                &   74.11(4) &  75.51(15) &   61.24(7) &   62.22(6)  &   60.72(4) &   45.6( 6) &   45.1(1.1) &    44.7(9)  \\
$L_{1}/L_{total}(\%)$        &    40.8(1) &   40.6( 3) &    33.5(1) &    47.1(2)  &    44.5(1) &   35.1( 3) &    39.3( 9) &    34.5(6)  \\
$L_{3}/L_{total}(\%)$        &     ---    &    6.5( 6) &     ---    &     ---     &     ---    &    ---     &     ---     &     ---     \\
$\Omega_{1}=\Omega_{2}$      &   4.953(1) &  4.662( 8) &   6.987(3) &   4.273(4)  &   4.902(2) &  6.578(11) &   6.010(25) &   6.707(25) \\
$f(\%)$                      &     2.3(2) &    5.1(1.4)&    24.6(5) &    13.4(6)  &    14.5(4) &    6.2(1.8)&     8.0(4.2)&    14.9(4.0)\\
$r_{1}$(pole)                &  0.3093(1) & 0.3198( 4) &  0.2699(2) &  0.3375(2)  &  0.3154(2) & 0.2715( 8) &  0.2837(1)  &  0.2719(16) \\
$r_{1}$(side)                &  0.3232(1) & 0.3347( 5) &  0.2825(2) &  0.3546(2)  &  0.3306(2) & 0.2832( 9) &  0.2963(1)  &  0.2840(19) \\
$r_{1}$(back)                &  0.3563(2) & 0.3686( 6) &  0.3240(4) &  0.3913(3)  &  0.3681(2) & 0.3179(15) &  0.3316(2)  &  0.3217(34) \\
$r_{2}$(pole)                &  0.4067(1) & 0.3990(10) &  0.4670(2) &  0.3914(6)  &  0.4139(3) & 0.4500( 7) &  0.4389(2)  &  0.4571(19) \\
$r_{2}$(side)                &  0.4312(1) & 0.4226(13) &  0.5049(3) &  0.4145(7)  &  0.4404(3) & 0.4830(10) &  0.4696(3)  &  0.4921(26) \\
$r_{2}$(back)                &  0.4602(2) & 0.4529(19) &  0.5332(3) &  0.4480(11) &  0.4722(5) & 0.5096(12) &  0.4975(4)  &  0.5197(34) \\
$\theta_{s}(^{\circ})$       &    ---     &   ---      &   ---      &  6$^{t}$    &  97$^{t}$  &   ---      &  86$^{t}$   &  92$^{t}$   \\
$\psi_{s}(^{\circ})$         &    ---     &   ---      &   ---      &     198( 1) &     312(2) &   ---      &     279( 8) &     295( 4) \\
$r_{s}(^{\circ})$            &    ---     &   ---      &   ---      &    46.4( 9) &     8.8(2) &   ---      &    14.4( 5) &    15.1( 3) \\
$T_{s}/T_{\ast}$             &    ---     &   ---      &   ---      &  1.18$^{t}$ & 0.60$^{t}$ &   ---      &  0.90$^{t}$ &  0.70$^{t}$ \\
on Star                      &    ---     &   ---      &   ---      &      1      &     2      &   ---      &    2        &    2        \\
$\overline\Sigma\times 10^4$ &  1.6021    & 1.5806     &  0.6722    &  0.4043     &  0.4040    & 0.2618     &  0.2112     &  0.1747     \\
\hline
\end{tabular}
\end{table*}

\begin{table}[htbp]
\caption{Photometric solutions for the multi-band light curves of V1038 Her.}\label{tabwd3}
\centering          
\begin{tabular}{ll}
\hline\hline
Parameters                   &  Values     \\
\hline 
$q$ ($M_2/M_1$)              &  1.544(6)   \\
$\Omega_{in}$                & 4.5914      \\
$\Omega_{out}$               & 4.0118      \\
$T_2-T_1$ (K)                &   -138(5)   \\
$i(^{\circ})$                &  76.85(19)  \\
$L_1/L_{total}$ (B)(\%)     & 45.0(4)     \\
$L_1/L_{total}$ (V)(\%)     & 42.7(4)     \\
$L_1/L_{total}$ (Ic)(\%)     & 38.7(4)     \\
$L_3/L_{total}$ (B)(\%)     & 1.7(13)     \\
$L_3/L_{total}$ (V)(\%)     & 4.4(10)     \\
$L_3/L_{total}$ (Ic)(\%)     & 10.2(7)     \\
$\Omega_{1}=\Omega_{2}$      & 4.558(7)    \\
$f(\%)$                      &   5.8(12)   \\
$\overline\Sigma\times 10^4$ & 1.1954      \\
\hline 
\end{tabular}
\end{table}

\begin{table*}[htbp]
\caption{The Spectroscopy log from APOGEE and the newly determined velocities.}
\label{tabsp1}
\scriptsize
\begin{center}
\begin{tabular}{ccccccrrrr}
\hline\hline  
Targets & UT-MID$^{\ast}$  & HJD$^{\ast}$ & EXPTIME$^{\ast}$ & SNR$^{\ast}$ &  Phase$^{\dagger}$ & \multicolumn{1}{c}{$V_{p}$}  & \multicolumn{1}{c}{$V_{s}$} & \multicolumn{1}{c}{$V_3$} \\
         &                       &  (2,400,000+) &   (s)    &     &        &   \multicolumn{1}{c}{(km/s)} & \multicolumn{1}{c}{(km/s)}  & \multicolumn{1}{c}{(km/s)} \\\hline
RV CVn    & 2011-05-22T06:04:56.8 & 55703.756507 & 3194.70 &  57 &  0.008 &   14.2$\pm$1.3 &     ---         & --- \\
          & 2011-05-23T07:05:44.0 & 55704.798654 & 5324.50 &  84 &  0.874 &  -88.2$\pm$3.5 &   205.5$\pm$6.5 & --- \\
          & 2011-07-17T03:33:02.9 & 55759.646863 & 4792.05 &  82 &  0.342 &  118.8$\pm$2.2 &  -191.3$\pm$2.5 & --- \\
          & 2012-04-12T09:55:22.1 & 56029.918053 & 2002.04 &  36 &  0.954 &  -49.0$\pm$2.0 &   163.7$\pm$3.2 & --- \\
          & 2012-05-02T07:14:41.4 & 56049.805913 & 2002.04 &  50 &  0.731 & -117.1$\pm$1.1 &   240.6$\pm$1.4 & --- \\
          & 2013-01-23T11:55:18.5 & 56315.998511 & 3503.56 &  63 &  0.213 &  140.7$\pm$1.5 &  -193.6$\pm$1.5 & --- \\
          & 2018-03-04T09:00:57.2 & 58181.879788 & 2002.04 &  49 &  0.001 &    7.2$\pm$1.2 &         ---     & --- \\
          & 2018-04-04T08:24:35.9 & 58212.855093 & 2502.54 &  63 &  0.909 &  -62.8$\pm$2.5 &  183.61$\pm$6.3 & --- \\
          & 2018-04-25T06:15:24.3 & 58233.765016 & 3003.05 &  48 &  0.477 &        ---     &    -2.9$\pm$1.5 & --- \\
          & 2018-05-02T05:20:33.1 & 58240.726668 & 2002.04 &  63 &  0.302 &  135.7$\pm$1.6 &  -202.0$\pm$1.9 & --- \\
          & 2018-05-05T07:25:17.4 & 58243.813170 & 2002.04 &  59 &  0.752 & -111.1$\pm$1.3 &   245.8$\pm$1.8 & --- \\
          & 2018-05-24T04:20:28.2 & 58262.683816 & 1001.02 &  28 &  0.756 & -109.4$\pm$1.1 &   261.1$\pm$1.6 & --- \\\hline
EK Com    & 2016-04-27T04:04:04.6 & 57505.673517 & 4025.47 & 127 &  0.910 &  -45.6$\pm$1.0 &         ---     & --- \\
          & 2017-05-15T04:43:42.1 & 57888.699969 & 1341.82 &  76 &  0.162 &   50.7$\pm$1.7 &  -284.9$\pm$2.5 & --- \\
          & 2017-05-16T03:58:38.3 & 57889.668602 & 3631.43 & 127 &  0.792 &  -95.3$\pm$3.0 &   218.3$\pm$6.9 & --- \\\hline
V384 Ser  & 2018-06-11T07:05:46.6 & 58280.799349 & 4025.47 & 176 &  0.324 &   85.7$\pm$5.6 &  -200.2$\pm$7.4 & -2.1$\pm$0.2 \\
          & 2018-06-12T03:40:01.1 & 58281.656430 & 1842.33 & 122 &  0.513 &  -21.0$\pm$1.0 &         ---     & -2.3$\pm$0.3 \\\hline
V1038 Her & 2016-05-07T10:41:58.1 & 57515.948889 & 1341.82 & 104 &  0.646 & -126.1$\pm$1.4 &  180.3$\pm$1.8  & -2.6$\pm$0.5 \\
          & 2016-05-08T07:41:14.9 & 57516.823405 & 1341.82 &  89 &  0.907 &  -61.3$\pm$4.9 &  121.3$\pm$5.1  & -1.2$\pm$0.4 \\
          & 2016-05-09T10:17:35.2 & 57517.931995 & 1341.82 &  89 &  0.041 &   -3.7$\pm$0.9 &    ---          & -1.8$\pm$0.6 \\
          & 2017-06-15T05:50:09.9 & 57919.746312 &  447.27 &  66 &  0.342 &  138.2$\pm$1.5 & -171.9$\pm$1.5  &  2.0$\pm$0.4 \\
          & 2017-06-16T04:22:22.8 & 57920.685336 & 4579.21 & 186 &  0.843 &  -93.1$\pm$3.9 &  142.6$\pm$5.2  & -1.4$\pm$0.4 \\\hline
EH CVn    & 2017-04-17T05:39:11.7 & 57860.739841 & 2683.64 &  69 &  0.441 &  -35.0$\pm$0.8 &    ---          &  ---         \\
          & 2017-04-18T05:14:02.7 & 57861.722358 & 1341.82 &  52 &  0.169 &   63.5$\pm$1.0 &--229.1$\pm$1.2  &  ---         \\\hline
J125403   & 2018-03-05T10:12:23.4 & 58182.928935 & 2002.04 &  41 &  0.784 &  -57.7$\pm$1.3 &  166.4$\pm$2.3  &  ---         \\
          & 2018-03-09T10:09:39.3 & 58186.927029 & 2502.54 &  36 &  0.658 &  -26.6$\pm$1.2 &    ---          &  ---         \\
          & 2018-03-31T07:50:41.8 & 58208.830191 & 2002.04 &  12 &  0.144 &    4.5$\pm$3.6 &    ---          &  ---         \\
          & 2018-04-03T08:42:30.7 & 58211.866085 & 2002.04 &  30 &  0.438 &  -12.4$\pm$0.7 &    ---          &  ---         \\
          & 2018-04-07T08:10:23.2 & 58215.843660 & 3003.05 &  45 &  0.236 &   29.1$\pm$1.7 & -170.6$\pm$2.4  &  ---         \\
          & 2018-04-23T06:39:31.7 & 58231.779918 & 1001.02 &  17 &  0.523 &  -17.2$\pm$1.3 &    ---          &  ---         \\\hline
\end{tabular}
\end{center}
\begin{flushleft}
\textbf{Note.} $^{\ast}$Retrieved from apVisit files. $^{\dagger}$Determined using the following initial ephemerides (HJD) and periods: \\
RV CVn: 2456001.8955 (for MJD 55703 to 56315; \citealt{Diethelm12}) and 2457855.4339 (for MJD 58000+; \citealt{Pagel18}), 0.269567 d; \\
EK Com: 2457465.5615 \citep{Hubscher17}, 0.266685 d; V384 Ser: 2458290.3866 \citep{Pagel20}, 0.268729 d; 
V1038 Her: 2457125.57364 \citep{Jurysek17}, 0.268180 d; EH CVn: 2458172.1786 \citep{Xiaq18}, 0.263583 d; J125403: 2458743.16197 (this study), 0.268798 d. The phase for EK Com is half cycle corrected from \citet{Hubscher17} to be in phase with its light curve. 
\end{flushleft}
\end{table*}

\begin{table*}[htbp]
\caption{The spectroscopy log from LAMOST and the newly determined velocities.}\label{tabsp2}
\scriptsize
\begin{center}
\begin{tabular}{clccccccccc}
\hline\hline
Targets & DATE-OBS$^{\ast}$ & HJD$^{\ast}$ & EXPTIME$^{\ast}$ & SNR$^{\ast}$ &  phase$^{\dagger}$ & $V_{p}$   &  $V_{s}$ & $V_3$  \\
        &      & (2,400,000+) &  (s)   & B band &      & (km/s)  & (km/s) & (km/s) \\\hline
V384Ser & 2020-06-07T14:27:27 & 59008.106167 & 1200 & 17 & 0.766 & -126.3$\pm$3.9 &  197.0$\pm$2.8 & -14.9$\pm$2.0 \\
        & 2020-06-07T14:50:48 & 59008.122381 & 1200 & 16 & 0.826 & -123.2$\pm$3.1 &  177.2$\pm$2.3 & -10.7$\pm$1.3 \\
        & 2020-06-07T15:14:12 & 59008.138631 & 1200 & 14 & 0.887 & -107.8$\pm$5.0 &  128.4$\pm$4.3 &  -3.6$\pm$2.4 \\\hline
EHCVn   & 2019-03-13T18:08:55 & 58556.260475 & 1200 & 16 & 0.157 &   52.8$\pm$1.6 & -200.1$\pm$1.6 &  ---         \\
        & 2019-03-13T18:32:14 & 58556.276668 & 1200 & 15 & 0.219 &   70.1$\pm$2.1 & -220.8$\pm$2.4 &  ---         \\
        & 2019-03-13T18:55:33 & 58556.292861 & 1200 & 14 & 0.280 &   62.4$\pm$1.4 & -228.4$\pm$1.7 &  ---         \\
        & 2020-04-11T16:23:19 & 58951.187253 & 1200 & 28 & 0.459 &    ---         &  -42.9$\pm$0.3 &  ---         \\
        & 2020-04-11T16:46:41 & 58951.203479 & 1200 & 25 & 0.520 &    ---         &  -38.7$\pm$0.7 &  ---         \\
        & 2020-04-11T17:10:04 & 58951.219717 & 1200 & 29 & 0.582 &  -41.6$\pm$0.8 &    ---         &  ---         \\
\hline
\end{tabular}
\end{center}
\begin{flushleft}
\textbf{Note.} $^{\ast}$Retrieved from single exposure spectra. $^{\dagger}$Determined using the following initial ephemerides (HJD) and periods: \\
V384 Ser: 2458985.59583 (this study), 0.268729 d; EH CVn: 2458172.1786 \citep{Xiaq18}, 0.263583 d
\end{flushleft}
\end{table*}

\begin{table*}[htbp]
\caption{The observation of EK Com with MRES and the newly determined velocities.}\label{tabsp3}
\scriptsize
\begin{center}
\begin{tabular}{cccccccccc}
\hline\hline
 DATE-OBS$^{\ast}$  & HJD & EXPTIME &  Phase$^{\dagger}$ & \multicolumn{1}{c}{$V_{p}$}  & \multicolumn{1}{c}{$V_{s}$} \\
                    &  (2,400,000+) & (s) &        &   \multicolumn{1}{c}{(km/s)} & \multicolumn{1}{c}{(km/s)} \\\hline
2017-04-05T18:21:13 &  57849.26955 &  900 &  0.307 &  33.3$\pm$ 1.5 & -274.8$\pm$ 1.5 \\
2017-04-05T18:38:41 &  57849.28168 &  900 &  0.352 &  18.2$\pm$ 1.3 & -247.7$\pm$ 1.4 \\
2017-04-05T19:42:35 &  57849.32605 &  900 &  0.519 & -27.2$\pm$ 0.8 &    ---          \\
2017-04-05T19:58:01 &  57849.33677 &  900 &  0.559 &      ---       &    ---          \\
2017-04-05T21:06:11 &  57849.38411 &  900 &  0.737 & -78.5$\pm$ 1.7 &  249.0$\pm$ 1.7 \\
2017-04-05T21:21:37 &  57849.39483 &  900 &  0.777 & -79.6$\pm$ 1.7 &  241.5$\pm$ 1.7 \\
\hline
\end{tabular}
\end{center}
\end{table*}

\begin{table*}[htbp]
\caption{The radial velocity fits and absolute elements of targets.}\label{tababs}
\scriptsize
\centering
\begin{tabular}{lccccccccccc}
\hline\hline
Objects  & $V_0$ (km/s) &  A & \scriptsize{$(M_1+M_2)\sin^3 i$}  & $M_1$  & $M_2$  & $R_1$  & $R_2$  & $L_2$  & $L_{tot}$ & $L_{G2}$ \\\hline
RV CVn    &  12(3) & 1.98(4)  &1.41(8)  & 0.48(3)  & 0.95(6)  & 0.65(3) & 0.89(3) & 0.34(1) & 0.55(2) & 0.564(11) \\
EK Com    & -30(9) & 1.90(10) &1.28(19) & 0.28(4)  & 1.02(15) & 0.53(5) & 0.95(9) & 0.52(5) & 0.73(7) & 0.485(10) \\
V384 Ser  & -14.7(5) & 1.92(5) &1.09(7)& 0.48(3)  & 0.84(5)  & 0.65(3) & 0.83(3) & 0.31(1) & $^{\ast}$0.61(2) & 0.846(13) \\
EH CVn    & -39(4) & 2.00(6)  &1.02(6)  & 0.55(5)  & 1.00(8)  & 0.68(2) & 0.89(3) & 0.36(2) & 0.64(4) & 0.484(8) \\
J125403   & -11(6) & 1.71(10) &0.32(5)  & 0.23(4)  & 0.71(13) & 0.50(3) & 0.84(5) & 0.32(2) & 0.49(4) & 0.761(19) \\
V1038 Her &   4(9) & 1.95(15) &1.26(26) & 0.53(12) & 0.86(20) & 0.67(5) & 0.83(6) & 0.31(5) & $^{\ast}$0.59(9) & 0.623(31) \\
\hline
\end{tabular}
\begin{flushleft}
\textbf{Note.} Parameters without units noted are in the units of solar quantities.
$^{\ast}$Including the luminosity of a third body.
\end{flushleft}
\end{table*}

\begin{table*}[htbp]
\scriptsize
\caption{Parameters of K-type CBs with both photometric and spectroscopic measurements}\label{tabcollect}
\centering
\begin{tabular}{llllccllcllcl}
\hline\hline
Name          & Period  & $q_{sp}$ & $M\sin^3 i$ & $i$ & $M_p$  & $M_s$   & $T_p$  & $T_{p}-T_{s}$ & $f$  &  eL3  & A/W  & Refs    \\
              & (days)  &        & (M$_{\odot}$) & ($^{\circ}$)& (M$_{\odot}$) &  (M$_{\odot}$)& (K)  & (K) & (\%)  &    &    &  \\
\hline
   CC Com & 0.2207  & 0.527  & 1.083  &  89.8   & 0.717    & 0.377   & 4200 & -100$\pm$60  & 17     & ...      & W  & (1),(2)    \\
J1601     & 0.2265  & 0.67   & ...    &  79.5   & 0.86     & 0.57    & 4500 &  ...         & 10     & ...      & W  & (3)        \\
 V523 Cas & 0.2337  & 0.516  & 1.110  &  83.1   & 0.75     & 0.38    & 4410 & -326$\pm$5   &16$\ast$& ...      & W  & (4),VIII   \\
   RW Com & 0.2373  & 0.471  & 1.052  &  74.9   & 0.80     & 0.38    & 4720 & -180$\pm$20  & 6.1    & ...      & W  & (5),XIV    \\
J1508     & 0.2601  & 0.51   & ...    &  90     & 1.07     & 0.55    & 4500 &  ...         & 12     & ...      & W  & (3)        \\
   IL Cnc & 0.2677  & 0.57   & ...    &  73.6   & 0.90     & 0.51    & 4720 & -280$\pm$20  &  9     & ...      & W  & (6)        \\
    i Boo & 0.2678  & 0.487  & 1.132  &  72.8   & 0.98     & 0.55    & 5300 &  265$^a$     & ...    & 0.40$^c$ & W (A)& (7),IV     \\
V1167 Her & 0.2753  & 0.662  & 1.050  &  71.6   & 0.74     & 0.49    & 4640 & -260$\pm$30  & 9.5    & 0.48 V   & W  & (5), XIII  \\
   VW Cep & 0.2783  & 0.272  & 0.85   &  65.0   & 0.897    & 0.247   & 4930 & -270$\pm$100 & ...    & 0.12$^c$ & W  & (8),(9)    \\
   XY Leo & 0.2841  & 0.729  & 1.188  &  71.1   & 0.813    & 0.593   & 4524 & -326$\pm$10  & 8      & 0.13     & W  & (10),(11),XII \\
   RW Dor & 0.2854  & 0.68   & 0.987  &  76.9   & 0.64     & 0.43    & 4780 & -420$\pm$200 & --     & ...      & W  & (12),(13)  \\
   PY Vir & 0.3113  & 0.773  & 1.387  &  68.9   & 0.964    & 0.745   & 4559 & -271$\pm$136 & 0      & 0.08     & W  & (14),VIII  \\
   BH Cas & 0.4059  & 0.475  & ...    &  71.7   & 0.73$^e$ & 0.35    & 5555 & -445$\pm$22  & 22     & ...      & W  & (15),(16)  \\
   AH Vir & 0.4075  & 0.303  & 1.754  &  85.2   & 1.360    & 0.412   & 5300 & -480$\pm$200 & 23     & Y        & W  & (17)       \\\hline
   RV CVn & 0.2696  & 0.507  & 1.41   &  84.2   & 0.95     & 0.48    & 4703 & -147$\pm$2   &  9.8   &$<$ 0.01 T& W  & This study \\
   EK Com & 0.2667  & 0.275  & 1.28   &  84.2   & 1.02     & 0.28    & 5000 & -354$\pm$8   & 13.1   &$<$ 0.01 T& W  & This study \\
 V384 Ser & 0.2687  & 0.575  & 1.09   &  69.6   & 0.84     & 0.48    & 4750 & -312$\pm$31  &  5.4   &  0.09 T  & W  & This study \\
   EH CVn & 0.2636  & 0.551  & 1.02   &  60.7   & 1.00     & 0.55    & 4750 & -365$\pm$6   & 14.5   &  ...     & W  & This study \\
J125403   & 0.2688  & 0.319  & 0.32   &  44.7   & 0.71     & 0.23    & 4750 & -454$\pm$55  & 14.9   &  ...     & W  & This study \\
V1038 Her & 0.2682  & 0.620  & 1.26   &  75.5   & 0.86     & 0.53    & 4750 & -192$\pm$3   &  5.1   & 0.065 T  & W  & This study \\\hline
   OT Cnc & 0.2178  & 0.474  & 0.280  &  37.6   & 0.835    & 0.396   & 4500 &   55$\pm$40  & 43     & 0.365 V  & A/W& (18),(19)  \\
  J0930B  & 0.2277  & 0.397  & ...    &  86     & 0.86     & 0.341   & 4700 &  ...         & 17     & Y        & A/W& (20),(21)  \\
   VZ Psc & 0.2612  & 0.80   & 0.601  &  48     & 0.81     & 0.65    & 4500 &  390$\pm$60  & -5     & ...      & EB & (22),(23)  \\
 V345 Gem & 0.2748  & 0.142  & 1.054  &  61.2   & 1.371    & 0.195   & 6200 &  360$\pm$54  & 11     & 0.619 V  & A  & (19)       \\

\hline
\end{tabular}
\begin{flushleft}
\textbf{Note.} J1601 = 1SWASP J160156.04+202821.6; J1508 = 1SWASP J150822.80-054236.9; J0930B = 1SWASP J093010.78+533859.5 B \\
$^c$calculated. $^e$estimated. $^a$averaged value. T: Tess band. Y: a visual companion nearby ($< 2^{\prime\prime}$) \\
Refs:
(1) \cite{Kose11}, (2) \cite{Zola10}, (3) \cite{Lohr14}, (4) \cite{Zhang04}, (5) \cite{Djura11}, (6) \cite{LiuN20}, (7) \cite{Hill89a}, (8) \cite{Hill89b}, (9) \cite{Kaszas98}, (10) \cite{Djura06}, (11) \cite{Zola10}, (12) \cite{Hilditch92}, (13) \cite{Marino07}, (14) \cite{Deb11}, (15) \cite{Zola01}, (16) \cite{Metcalfe99}, (17) \cite{Lu93}, (18) \cite{Rucinski08a}, (19) \cite{Gazeas21}, (20) \cite{Lohr15}, (21) \cite{Koo14}, (22) \cite{Hrivnak95}, (23) \cite{Maceroni90}, VIII \cite{Rucinski03}, XIV \cite{Pribulla09b}, IV \cite{Lu01}, XIII \cite{Rucinski08b}, XII \cite{Pribulla07}
\end{flushleft}
\end{table*}

\begin{table}
\caption{Newly determined minimum light times.}\label{tabmin}
\centering
\begin{tabular}{cccc}
\hline\hline
BJD & Err & NA & Object \\
2,400,000+ & (days) &  &  \\
\hline
   58929.24999 &  0.00026 &  17 & RV CVn \\
   58929.92380 &  0.00009 &  16 & RV CVn \\
   58930.05881 &  0.00018 &  19 & RV CVn \\
   58930.46299 &  0.00007 &  15 & RV CVn \\
   58931.00221 &  0.00016 &  19 & RV CVn \\
   58931.13720 &  0.00017 &  19 & RV CVn \\
    ...        &  ...     &  ... & ...   \\
   58929.26165 &  0.00006 &  16 & EK Com \\
   58929.39510 &  0.00009 &  16 & EK Com \\
   58929.92845 &  0.00009 &  18 & EK Com \\
    ...        &  ...     &  ... & ...   \\
   58958.05189 &  0.00009 &  16 & V384 Ser \\
   58959.12703 &  0.00019 &  15 & V384 Ser \\
   58959.26099 &  0.00011 &  15 & V384 Ser \\
    ...        &  ...     &  ... & ...   \\
    ...        &  ...     &  ... & ...   \\
\hline
\end{tabular}
\end{table}

\end{document}